\documentclass[superscriptaddress,twocolumn,pre,nofootinbib]{revtex4-1}

\usepackage{amsmath}
\usepackage{graphicx,color}

\newcommand{\be}{\begin{equation}}
\newcommand{\ee}{\end{equation}}
\newcommand{\bea}{\begin{eqnarray}}
\newcommand{\eea}{\end{eqnarray}}

\begin{document}
\sloppy


\title{Models of universe with a polytropic equation of state: \\
III. The phantom universe}

\author{Pierre-Henri Chavanis}
\affiliation{Laboratoire de Physique Th\'eorique (IRSAMC), CNRS and UPS, Universit\'e de Toulouse, France}

\begin{abstract}

We construct models of universe with a generalized equation of state
$p=(\alpha \rho+k\rho^{1+1/n})c^2$ having a linear component and a
polytropic component. The linear equation of state $p=\alpha\rho c^2$
with $-1\le \alpha\le 1$ describes radiation ($\alpha=1/3$),
pressureless matter ($\alpha=0$), stiff matter ($\alpha=1$), and
vacuum energy ($\alpha=-1$). The polytropic equation of state
$p=k\rho^{1+1/n} c^2$ may be due to Bose-Einstein condensates with
repulsive ($k>0$) or attractive ($k<0$) self-interaction, or have
another origin. In this paper, we consider the case where the density
increases as the universe expands. This corresponds to a ``phantom
universe'' for which $w=p/\rho c^2<-1$ (this requires $k<0$). We
complete previous investigations on this problem and analyze in detail
the different possibilities. We describe the singularities using the
classification of [S. Nojiri, S.D. Odintsov, S. Tsujikawa,
Phys. Rev. D {\bf 71}, 063004 (2005)]. We show that for $\alpha>-1$
there is no Big Rip singularity although $w\le -1$. For $n=-1$, we
provide an analytical model of phantom bouncing universe
``disappearing'' at $t=0$.  We also determine the potential of the
phantom scalar field and phantom tachyon field corresponding to the
generalized equation of state $p=(\alpha \rho+k\rho^{1+1/n})c^2$.

\end{abstract}

\maketitle

\section{Introduction}

In previous papers of this series, we have constructed models of universe with a generalized equation of state
\begin{equation}
\label{intro1}
p=(\alpha \rho+k\rho^{1+1/n}) c^2,
\end{equation}
having a linear component and a polytropic component.  In Papers I and
II, we have assumed $\alpha+1+k\rho^{1/n}\ge 0$ corresponding to
$w=p/\rho c^2\ge -1$. In that case, the density decreases as the
universe expands. For $n>0$, the polytropic component dominates the
linear component when the density is high: This describes the early
universe (Paper I). For $n<0$, the polytropic component dominates the
linear component when the density is low: This describes the late
universe (Paper II).  When the polytropic pressure is positive
($k>0$), the solutions of the Friedmann equations exhibit past or
future singularities (or peculiarities).  When the polytropic pressure
is negative ($k<0$), there is no singularity. Furthermore, the
polytropic equation of state implies the existence of an upper bound
$\rho_{max}$ (in the past) and a lower bound $\rho_{min}$ (in the
future) for the density. It makes sense to identify the maximum
density to the Planck density $\rho_P=5.16 \, 10^{99}\, {\rm g}/{\rm
m}^3$ and the minimum density to the cosmological density
$\rho_{\Lambda}=7.02\, 10^{-24}\, {\rm g}/{\rm m}^3$. These constant
densities imply in turn the existence of two phases of exponential
inflation, one in the early universe and one in the late
universe. During the inflation, the universe is accelerating. The
early inflation is necessary to solve notorious difficulties such as
the singularity problem, the flatness problem, and the horizon problem
\cite{guth,linde}. The late inflation is necessary to account for the
observed accelerating expansion of our universe \cite{novae} driven by
dark energy \cite{cst}. In that context, the equation of state
(\ref{intro1}) with $k<0$ and $n<0$ corresponds to the generalized
Chaplygin gas \cite{chaplygin} that has been proposed as a model for
dark energy. From the generalized polytropic equation of state
(\ref{intro1}), we have obtained a model of universe without
singularity that possesses striking ``symmetries'' between the past
and the future (aioniotic universe). This model, which could have been
obtained from a principle of ``simplicity'' without making any
observation, turns out to be strikingly consistent with what we know
of the real universe. It is consistent with the standard model
\cite{weinberg,bt} but refines it by removing the primordial
singularity (Big Bang). In this model, the Planck density and the
cosmological density are interpreted as two {\it fundamental} bounds
for the density determined by the Planck constant $\hbar$
(microphysics) and the cosmological constant $\Lambda$ (cosmophysics),
respectively. These bounds differ by $122$ orders of magnitudes, a
difference that appears to be quite natural instead of representing a
``problem'' \cite{weinbergcosmo}.

In this paper, we consider a case that has not been treated in our previous papers. This is the case where the density increases as the universe expands. Since the nature of dark energy is unknown, this situation cannot be rejected {\it a priori}. It corresponds to an equation of state parameter $w$ less than $-1$ which violates the null dominant energy condition. This is referred to as a ``phantom universe'' \cite{caldwell} because  when the equation of state with $w<-1$ is constructed in terms of a scalar field, the corresponding kinetic term has the wrong sign (negative kinetic energy). It represents therefore a phantom (ghost) scalar field (see reviews \cite{cst,saridakis}).

Actually, there is a rich recent literature on this situation (more
than one thousand papers are related to phantom dark energy) since
observations do not exclude the possibility that we live in a phantom
universe. Indeed, observational data indicate that the equation of
state parameter $w$ lies in a narrow strip around $w=-1$ possibly
being below this value \cite{observations}.  The models based on
phantom dark energy usually predict a future singularity in which the
scale factor, the energy density, and the pressure of the universe
become infinite in a finite time. This would lead to the death of the
universe in a singularity called ``Big Smash'' \cite{innes}, ``Big
Rip'' or ``Cosmic Doomsday'' \cite{caldwellprl}. Contrary to the ``Big
Crunch'', the universe is destroyed not by excessive contraction but
rather by excessive expansion.  In phantom cosmology, every
gravitationally bound system ({\it e.g.} the solar system, the Milky
Way, the local group, galaxy clusters) is dissociated before the
singularity \cite{caldwellprl,np}, and the black holes gradually lose
their mass and finally vanish \cite{pedro,thermophantom}. This
scenario allows the explicit calculation of the rest of the lifetime
of our universe. Actually, as we approach the singularity, the energy
scale may grow up to the Planck one, giving rise to a second quantum
gravity era. Eventually, quantum effects may moderate or even prevent
the singularity \cite{quantum}. Other aspects of phantom cosmology
have been studied in \cite{ghosts}.

There are many interesting recent works on the study of singularities. In particular,  Nojiri {\it et al.}  \cite{classification} considered an equation of state of the form $p=-\rho-f(\rho)$ and obtained a classification of finite-time future singularities (see complements in \cite{fernandez}). They are of four types:

$\bullet$ Type 0 (Big Bang or Big Crunch): For $t\rightarrow t_s$, $a\rightarrow 0$, $\rho\rightarrow +\infty$, and $|p|\rightarrow +\infty$.

$\bullet$ Type I (Big Rip): For $t\rightarrow t_s$, $a\rightarrow +\infty$, $\rho\rightarrow +\infty$, and $|p|\rightarrow +\infty$.

$\bullet$ Type II (sudden singularity): For $t\rightarrow t_s$, $a\rightarrow a_s$, $\rho\rightarrow \rho_s$, and $|p|\rightarrow +\infty$.

$\bullet$ Type III (Big Freeze): For $t\rightarrow t_s$, $a\rightarrow a_s$, $\rho\rightarrow +\infty$, and $|p|\rightarrow +\infty$.

$\bullet$ Type IV (generalized sudden singularity): For $t\rightarrow t_s$, $a\rightarrow a_s$, $\rho\rightarrow \rho_s$, $|p|\rightarrow p_s$, and higher derivatives of $H$ diverge\footnote{We shall not consider this type of singularities in this paper.}.

In this classification,  $t_s$, $a_s$, $\rho_s$, and $p_s$ are all finite constants ($a_s\neq 0$). Type 0 is the standard Big Bang or Big Crunch singularity arising in the original Friedmann models \cite{weinberg}. Type I is the Big Rip singularity which emerges from the phantom equation of state $p=\alpha\rho c^2$ with constant $\alpha<-1$ \cite{caldwell,caldwellprl}, and from the equation of state (\ref{intro1}) with $\alpha=-1$, $k<0$ and $n<-2$ \cite{stefancic}. Type II corresponds to the sudden future singularity found by Barrow \cite{barrow} at which $a$ and $\rho$ are finite but $p$ diverges. Type III, arising in the equation of state (\ref{intro1}) with $n>0$ \cite{stefancic,bigfreeze} differs from the sudden future singularity in the sense that $\rho$ diverges. Type IV appears in the model described in \cite{classification}.

It is important to stress that the phantom models with $w<-1$ do not necessarily lead to future singularities. For example, the equation of state  (\ref{intro1}) with $\alpha=-1$ and $-2\le n<0$ does not present future singularity \cite{stefancic}. However, since the scale factor and the density increase indefinitely, this has been called ``Little Rip'' \cite{littlerip}.

On the other hand, the models with $w>-1$ may lead to past or future singularities. For example, the new form of primordial singularity (for $n>0$ and $k>0$)  described in Secs. IV C, VI and in Appendix A   of Paper I corresponds to a past singularity of type III: The universe starts at $t=0$ with a finite scale factor and an infinite density. On the other hand, the future singularity (for $-1<n<0$ and $k>0$) described in Sec. IV C and in Appendix B  of Paper II corresponds to a singularity of type II: At a finite time $t_s$, the universe reaches a point at which the scale factor is finite, the density vanishes and the pressure is infinite.

In paper II, we have also introduced a notion of ``peculiarity''. This
is when the density vanishes $\rho=0$ while the scale factor has a
finite value $a_s$ (when $a_s=0$ we shall call it generalized
peculiarity). In that case, the universe is empty (in other works, it
``disappears''). Although there is no singularity, this situation is
very peculiar. However, since the nature of dark energy is unknown,
all possibilities should be contemplated.

In this paper, we provide an exhaustive study of the equation of state (\ref{intro1}) in the case $w\le -1$ (requiring $k<0$) for arbitrary $-1\le \alpha\le 1$ and $n$. This is a natural continuation of our previous works which assumed $w\ge -1$ (Papers I and II). Our paper also completes previous studies of the case $w\ge -1$ that considered $\alpha=-1$ \cite{stefancic} or  $\alpha=0$ \cite{bigfreeze}. An interesting result of our study is that the equation of state (\ref{intro1}) with $\alpha>-1$ does {\it not} present a Big Rip singularity although $w<-1$, contrary to the linear equation of state $p=\alpha\rho c^2$ with $\alpha<-1$ \cite{caldwellprl} or the equation of state (\ref{intro1}) with $\alpha=-1$ and $n<-2$ \cite{stefancic}. Another interesting result of our study is the construction of a bouncing phantom universe for $-2<n<0$. For  $-1<n<0$, this bouncing universe presents a past singularity of type II: At $t=0$, the the pressure is infinite while the scale factor has a finite value and the density vanishes. For $\alpha>-1$ and $n=-1$, corresponding to a constant negative pressure, the bouncing phantom universe admits a simple analytical expression.

The paper is organized as follows. In Sec. \ref{sec_basic}, we recall the basic equations of cosmology. In Secs. \ref{sec_ges} and \ref{sec_dark}, we study the generalized equation of state (\ref{intro1}) for any value of the parameters $-1<\alpha\le 1$, $k<0$ and $n$, assuming $w<-1$ (phantom cosmology). In Sec. \ref{sec_scalar}, we determine the potential of the phantom scalar field  and the potential of the  phantom tachyon field corresponding to the generalized equation of state (\ref{intro1}). In Appendix \ref{sec_eosgm}, we treat the case $\alpha=-1$. In Appendix \ref{sec_summary}, we summarize all the results obtained in our series of papers and analyze the different singularities in terms of the classification of \cite{classification}.

\section{Basic equations of cosmology}
\label{sec_basic}

We assume that the universe is isotropic and homogeneous at large scales and contains a uniform perfect fluid of energy density $\epsilon(t)=\rho(t) c^2$ and pressure $p(t)$. We also assume that the universe is flat in agreement with observations of the cosmic microwave background (CMB) \cite{bt}. Finally, in this paper, we ignore the cosmological constant ($\Lambda=0$). It that case, the Einstein equations reduce to
\begin{equation}
\label{basic1}
\frac{d\rho}{dt}+3\frac{\dot a}{a}\left (\rho+\frac{p}{c^2}\right )=0,
\end{equation}
\begin{equation}
\label{basic2}
\frac{\ddot a}{a}=-\frac{4\pi G}{3} \left (\rho+\frac{3p}{c^2}\right ),
\end{equation}
\begin{equation}
\label{basic2b}
H^2=\left (\frac{\dot a}{a}\right )^2=\frac{8\pi G}{3}\rho,
\end{equation}
where $a(t)$ is the scale factor (``radius'' of the universe) and $H=\dot a/a$ is the Hubble parameter. These are the well-known Friedmann equations describing a non-static universe \cite{weinberg}. The first equation can be viewed as an ``equation of continuity''. For a given barotropic equation of state $p=p(\rho)$, it determines the relation between the density and the scale factor. Then, the temporal evolution of the scale factor is given by Eq. (\ref{basic2b}).  Introducing the equation of state parameter $w=p/\rho c^2$, we see from Eq. (\ref{basic2}) that the universe is decelerating if $w>-1/3$ (strong energy condition) and accelerating if $w<-1/3$. On the other hand, according to Eq. (\ref{basic1}), the density decreases with the scale factor if $w>-1$ (null dominant energy condition) and increases with the scale factor if $w<-1$. In this last case, we are dealing with a ``phantom universe''.

We will also need the thermodynamical equation
\begin{equation}
\label{a3}
\frac{dp}{dT}=\frac{1}{T}(\rho c^2+p),
\end{equation}
which can be derived from the first principle of thermodynamics
\cite{weinberg}. For a given barotropic equation of state $p=p(\rho)$,
this equation can be integrated to obtain the relation $T=T(\rho)$
between the temperature and the density. It can be shown
\cite{weinberg} that the Friedmann equations conserve the entropy of
the universe
\begin{equation}
\label{t4}
S=\frac{a^3}{T}(p+\rho c^2).
\end{equation}
If we impose that the entropy is positive, we conclude from Eq. (\ref{t4}) that the temperature is positive when $w>-1$ while it is negative when $w<-1$. The fact that a phantom universe has a negative temperature was mentioned in \cite{thermophantom}. Negative temperatures arise in other domains of physics such as 2D turbulence \cite{onsager}.

The simplest model of phantom universe corresponds to the linear equation of state $p=\alpha\rho c^2$ with $\alpha<-1$ \cite{caldwell}. The continuity equation (\ref{basic1}) can be integrated into
\begin{equation}
\label{leos5}
\rho \propto a^{3|1+\alpha|}.
\end{equation}
Substituting Eq. (\ref{leos5}) in Eq. (\ref{basic2b}) and solving the resulting equation for $a(t)$, we find that the scale factor, the Hubble parameter and the density increase in time as \cite{cst}:
\begin{equation}
\label{leos6}
a\propto (t_s-t)^{-2/(3 |1+\alpha|)},
\end{equation}
\begin{equation}
\label{leos7}
H=\frac{\dot a}{a}=\frac{2}{3 |1+\alpha|}(t_s-t)^{-1},
\end{equation}
\begin{equation}
\label{leos8}
\rho=\frac{1}{6\pi G(1+\alpha)^2}(t_s-t)^{-2}.
\end{equation}
They all diverge at a finite time $t=t_s$. This is the ``Big Rip'' singularity \cite{caldwellprl}, which is a singularity of type I \cite{classification}. We also find from Eq. (\ref{a3}) that the temperature behaves as
\begin{equation}
\label{leos8b}
T \propto -\rho^{\alpha/(\alpha+1)}\propto -a^{3|\alpha|}\propto -(t_s-t)^{-2\alpha/(1+\alpha)}.
\end{equation}
The temperature becomes more and more negative as the universe expands, and it diverges when $t\rightarrow t_s$.

\section{Generalized equation of state with $w<-1$}
\label{sec_ges}

We consider a generalized equation of state of the form
\begin{equation}
\label{basic3}
p=(\alpha \rho+k\rho^{1+1/n}) c^2.
\end{equation}
This is the sum of a standard linear equation of state $p=\alpha\rho c^2$ and a polytropic equation of state $p=k\rho^{\gamma} c^2$, where $k$ is the polytropic constant and $\gamma=1+1/n$ is the polytropic index. Concerning the linear equation of state, we assume $-1\le \alpha\le 1$  (the case $\alpha=-1$ is treated specifically in Appendix \ref{sec_eosgm}). Concerning the polytropic equation of state, we remain very general, so that $k$ and $n$ can take arbitrary values. In papers I and II,  we assumed that $\alpha+1+k\rho^{1/n}\ge 0$, so that the density decreases with the scale factor ($w\ge -1$). In the present paper, we assume that $\alpha+1+k\rho^{1/n}\le 0$ (a necessary condition is $k<0$) so that the density increases with the scale factor ($w\le -1$). This corresponds to a ``phantom universe''.

\subsection{The density}
\label{sec_density}

For the equation of state (\ref{basic3}), the Friedmann equation (\ref{basic1}) becomes
\begin{equation}
\label{d0}
\frac{d\rho}{dt}+3\frac{\dot a}{a}\rho (1+\alpha+k\rho^{1/n})=0.
\end{equation}
Assuming $\alpha+1+k\rho^{1/n}\le 0$, this equation can be integrated into
\begin{equation}
\label{d1}
\rho=\frac{\rho_*}{\left\lbrack 1-(a/a_*)^{3(1+\alpha)/n}\right\rbrack^n},
\end{equation}
where  $\rho_*=\lbrack (\alpha+1)/|k|\rbrack^n$ and $a_*$ is a constant of integration.

\begin{figure}[!h]
\begin{center}
\includegraphics[clip,scale=0.3]{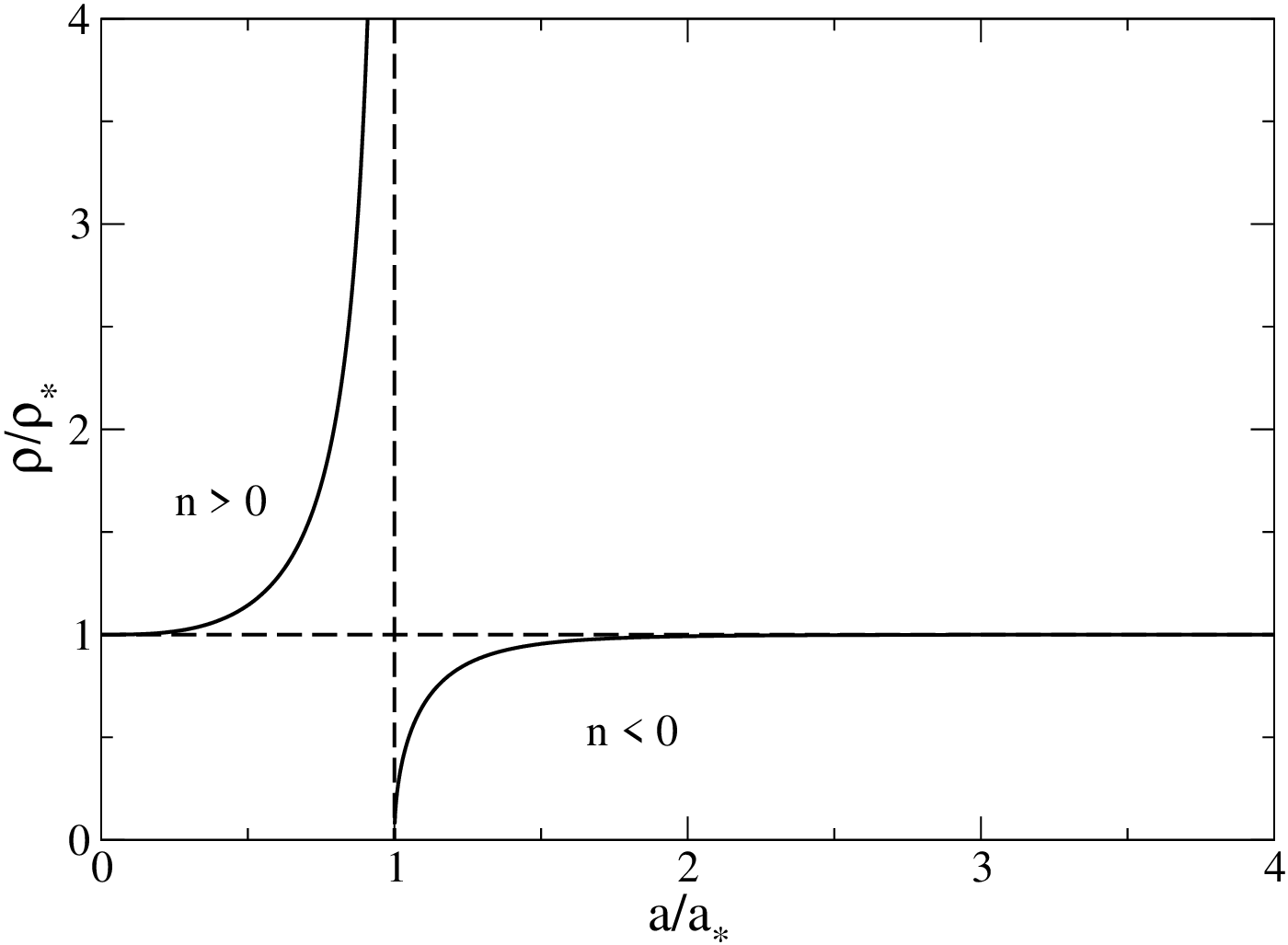}
\caption{Density as a function of the scale factor for $n>0$ and $n<0$ (specifically  $n=1$ and $n=-1/2$). We have taken $\alpha=0$.}
\label{densitenew}
\end{center}
\end{figure}

For $n>0$, the density is defined only when $a<a_*$. When $a\rightarrow 0$, $\rho\rightarrow \rho_*$ and $p\rightarrow -\rho_* c^2$. When $a\rightarrow a_*$,
\begin{equation}
\label{d2}
\frac{\rho}{\rho_*}\sim \left\lbrack \frac{n}{3(1+\alpha)}\right \rbrack^n \frac{1}{(1-a/a_*)^n}\rightarrow +\infty,
\end{equation}
and $p \rightarrow -\infty$.

For $n<0$, the density is defined only when $a>a_*$. When $a\rightarrow a_*$,
\begin{equation}
\label{d3}
\frac{\rho}{\rho_*}\sim \left\lbrack \frac{3(1+\alpha)}{|n|}\right \rbrack^{|n|} (a/a_*-1)^{|n|}\rightarrow 0.
\end{equation}
In the same limit,  $p \rightarrow -\infty$ for $n>-1$, $p$ tends to a finite value for $n=-1$, and $p\rightarrow 0$ for $n<-1$. On the other hand, when $a\rightarrow +\infty$, $\rho\rightarrow \rho_*$ and  $p\rightarrow -\rho_* c^2$.

Some curves giving the evolution of the density $\rho$  as a function of the scale factor $a$ are plotted in Fig. \ref{densitenew} for $n>0$ and $n<0$.

\subsection{The temperature}
\label{sec_gestemperature}

For the equation of state (\ref{basic3}), the thermodynamical equation (\ref{a3})  can be integrated into
\begin{equation}
\label{ges5}
T=-T_* \left \lbrack  ({\rho}/{\rho_*})^{1/n}-1\right \rbrack^{(\alpha+n+1)/(\alpha+1)}\left ({\rho}/{\rho_*}\right )^{\alpha/(\alpha+1)},
\end{equation}
where $T_*>0$ is a constant of integration. Combined with Eq. (\ref{d1}), we obtain
\begin{equation}
\label{ges6}
T=-T_* \frac{(a/a_*)^{3(\alpha+n+1)/n}}{\left \lbrack 1-(a/a_*)^{3(1+\alpha)/n}\right \rbrack^{n+1}}.
\end{equation}
We have to consider different cases.

We first assume $n>0$. When $a\rightarrow 0$, $T\rightarrow 0$; when $a\rightarrow a_*$, $T\rightarrow -\infty$.

We now assume $n<0$. When $a\rightarrow a_*$, $T\rightarrow 0$ for $n<-1$ and $T\rightarrow -\infty$ for $n>-1$. When $a\rightarrow +\infty$, $T\rightarrow 0$ for $n+\alpha+1>0$ and $T\rightarrow -\infty$ for $n+\alpha+1<0$.

\begin{figure}[!h]
\begin{center}
\includegraphics[clip,scale=0.3]{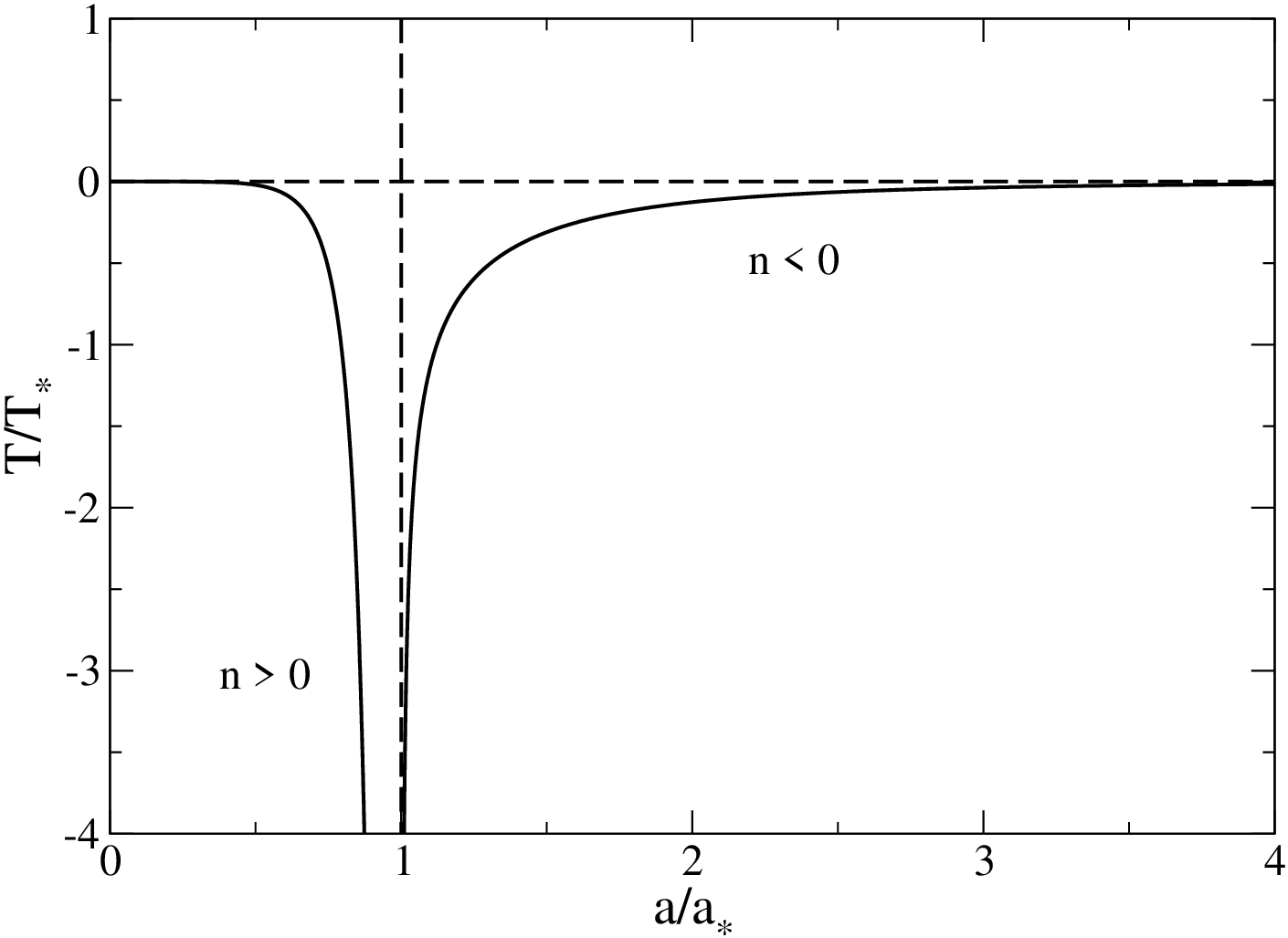}
\caption{Temperature as a function of the scale factor for $n>0$ and $n<0$ (specifically  $n=1$ and $n=-1/2$). We have taken $\alpha=0$.}
\label{tempnew}
\end{center}
\end{figure}

The extremum of temperature (when it exists) is located at
\begin{equation}
\label{s2}
\frac{\rho_e}{\rho_*}=\left \lbrack \frac{\alpha n}{(1+\alpha)(n+1)}\right \rbrack^n,
\end{equation}
\begin{equation}
\label{s3}
\frac{a_e}{a_*}=\left (-\frac{\alpha+n+1}{\alpha n}\right )^{n/\lbrack 3(1+\alpha)\rbrack},
\end{equation}
\begin{eqnarray}
\label{ss4}
\frac{T_{e}}{T_*}=-\left (-\frac{n+\alpha+1}{n\alpha}\right )^{\frac{n+\alpha+1}{1+\alpha}}\left\lbrack \frac{n\alpha}{(\alpha+1)(n+1)}\right\rbrack^{n+1}.\nonumber\\
\end{eqnarray}

Some curves giving the evolution of the temperature $T$  as a function of the scale factor $a$ are plotted in Fig. \ref{tempnew} for $n>0$ and $n<0$.

Finally, the entropy (\ref{t4}) is given by
\begin{equation}
\label{ges12}
S=(\alpha+1)\frac{a_*^3}{T_*}\rho_* c^2,
\end{equation}
and we explicitly check that it is a positive constant.

\subsection{The parameter $w(t)$}
\label{sec_w}

We can rewrite the equation of state (\ref{basic3}) as $p=w(t) \rho c^2$ with
\begin{equation}
\label{w2}
w(t)=\alpha- (\alpha+1)\left (\frac{\rho}{\rho_*}\right )^{1/n}.
\end{equation}

For $n>0$, $w\rightarrow -1$ when $a\rightarrow 0$ and $w\rightarrow -\infty$ when $a\rightarrow a_*$.

For $n<0$, $w\rightarrow -\infty$ when $a\rightarrow a_*$ and $w\rightarrow -1$ when $a\rightarrow +\infty$.

\begin{figure}[!h]
\begin{center}
\includegraphics[clip,scale=0.3]{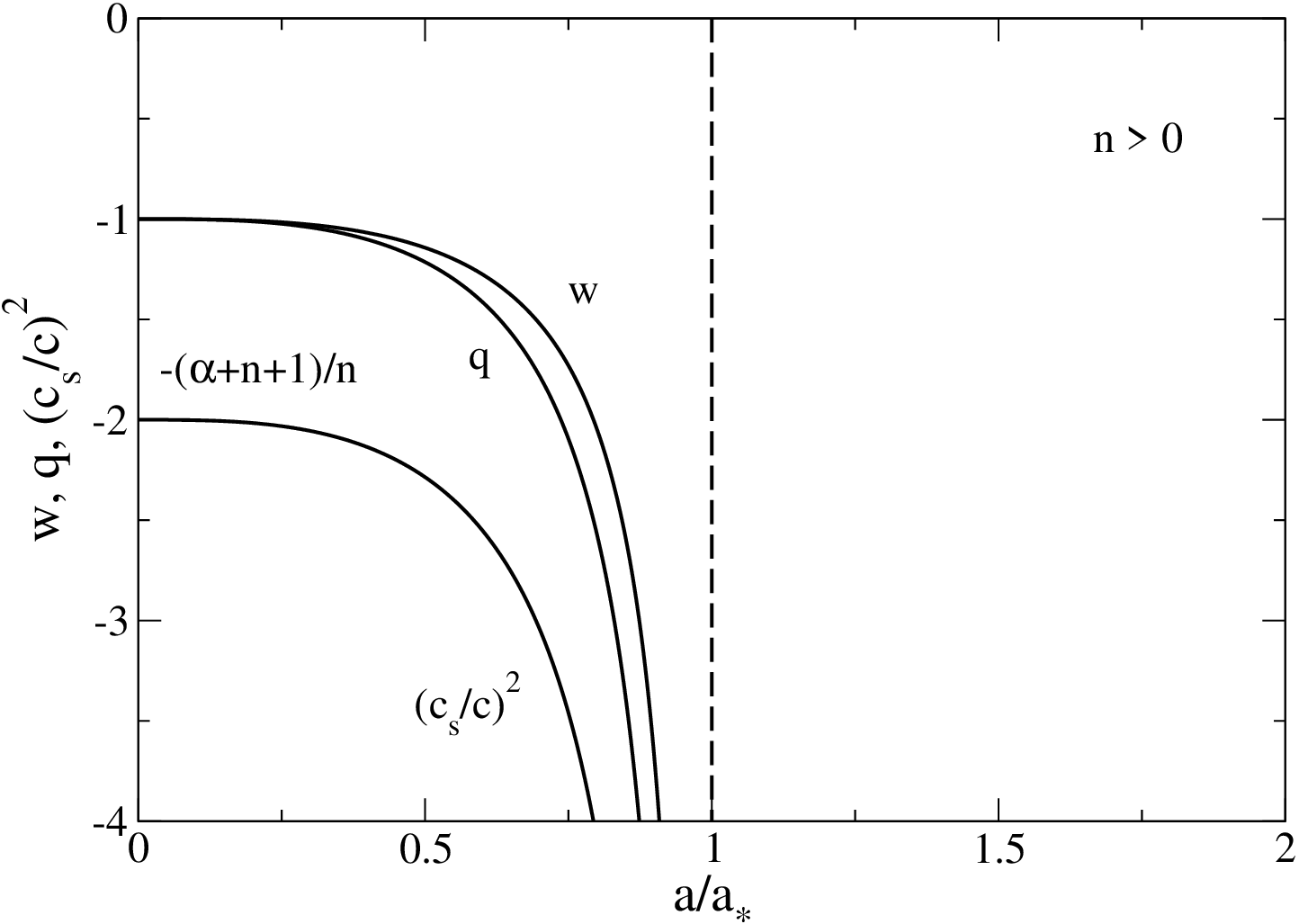}
\caption{The parameters $w$, $q$ and $c_s^2/c^2$ as a function of the scale factor $a$ for  $n>0$ (specifically  $n=1$). We have taken $\alpha=0$.}
\label{newwNpos}
\end{center}
\end{figure}

\begin{figure}[!h]
\begin{center}
\includegraphics[clip,scale=0.3]{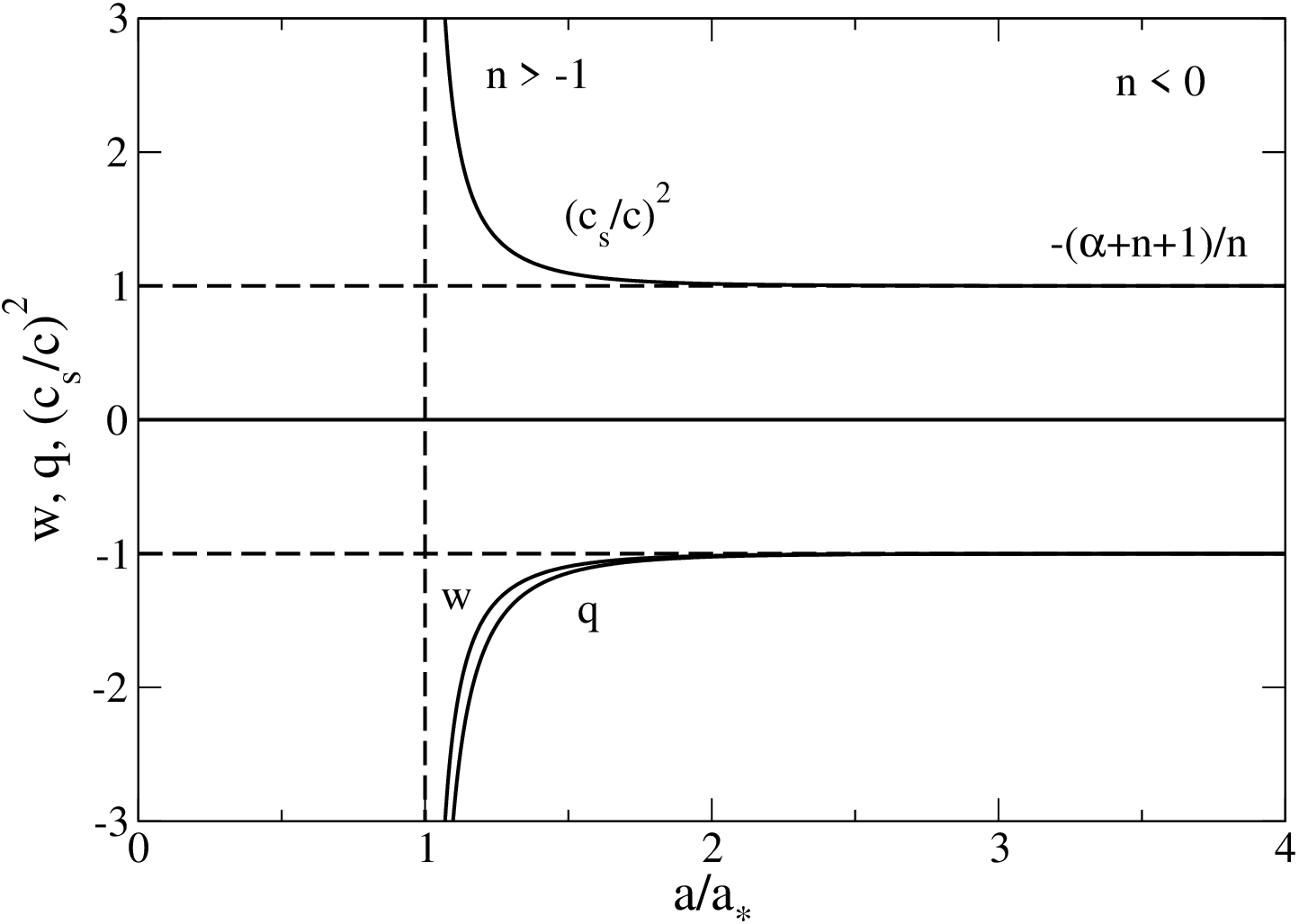}
\caption{The parameters $w$, $q$ and $c_s^2/c^2$ as a function of the scale factor $a$ for  $n<0$ (specifically $n=-1/2$). We have taken $\alpha=0$.}
\label{newwNneg}
\end{center}
\end{figure}

Some curves giving the evolution of  $w$  as a function of the scale factor $a$ are plotted in Figs. \ref{newwNpos} and \ref{newwNneg} for $n>0$ and $n<0$.

\subsection{The velocity of sound}
\label{sec_sound}

For the equation of state (\ref{basic3}), the velocity of sound is given by
\begin{equation}
\label{s1}
c_s^2=p'(\rho)= \left \lbrack \alpha- (\alpha+1)\frac{n+1}{n}\left (\frac{\rho}{\rho_*}\right )^{1/n}\right \rbrack c^2.
\end{equation}
For $n<0$, the velocity of sound vanishes at the point (\ref{s2})-(\ref{ss4}) where the temperature is extremum.  At that point, the pressure is maximum with value
\begin{equation}
\label{hjy1}
\frac{p_e}{\rho_*c^2}=\frac{\alpha}{n+1}\left \lbrack \frac{\alpha n}{(1+\alpha)(n+1)}\right \rbrack^n. \end{equation}
The case $c_s^2<0$ corresponds to an imaginary velocity of sound. We also define
\begin{equation}
\label{s4}
\frac{\rho_s}{\rho_*}=\left \lbrack -\frac{(1-\alpha)n}{(1+\alpha)(n+1)}\right \rbrack^n,
\end{equation}
\begin{equation}
\label{s5}
\frac{a_s}{a_*}=\left\lbrack \frac{\alpha+2n+1}{n(1-\alpha)}\right \rbrack^{n/\lbrack 3(1+\alpha)\rbrack},
\end{equation}
corresponding to a possible point where the velocity of sound is equal to the speed of light. Different cases have to be considered.

We first assume $n>0$. When $a\rightarrow 0$,  $(c_s/c)^2\rightarrow -(\alpha+n+1)/n$;  when $a\rightarrow a_*$, $(c_s/c)^2\rightarrow -\infty$. The velocity of sound is always imaginary.

We now assume $n<0$. When $a\rightarrow a_*$, $(c_s/c)^2\rightarrow +\infty$ for  $n>-1$ and $(c_s/c)^2\rightarrow -\infty$ for $n<-1$; when  $a\rightarrow +\infty$, $(c_s/c)^2\rightarrow -(\alpha+n+1)/n$. For $n>-1$ and  $\alpha+n+1>0$,  $c_s^2$  in always positive. For $n>-1$ and $\alpha+n+1<0$, it is positive for $a<a_e$ and negative for $a>a_e$. For $n<-1$ and $\alpha+n+1<0$, $c_s^2$  in always negative. For $n<-1$ and $\alpha+n+1>0$, it is negative for $a<a_e$ and positive for $a>a_e$.
For $n>-1$ and  $\alpha+2n+1>0$, the velocity of sound is always larger than the speed of light. For $n>-1$ and $\alpha+2n+1<0$,  the velocity of sound is larger than the speed of light for $a<a_s$ and smaller for $a>a_s$. For $n<-1$ and $\alpha+2n+1<0$, velocity of sound is always smaller than the speed of light. For $n<-1$ and $\alpha+2n+1>0$,  the velocity of sound is smaller than the speed of light for $a<a_s$ and larger for $a>a_s$.

Some curves giving the evolution of  $(c_s/c)^2$  as a function of
the scale factor $a$ are plotted in Figs. \ref{newwNpos} and \ref{newwNneg}
for $n>0$ and $n<0$.

\section{Evolution of the scale factor}
\label{sec_dark}

\subsection{The deceleration parameter}
\label{sec_dec}

The deceleration parameter is defined by Eqs. (I-77) and (I-78). A phantom universe is always accelerating since $q\le -1<0$. For the equation of state (\ref{basic3}), using Eq. (\ref{w2}), we get
\begin{equation}
\label{dark3}
q(t)=\frac{1+3\alpha}{2}- \frac{3}{2}(\alpha+1)\left (\frac{\rho}{\rho_*}\right )^{1/n}.
\end{equation}

For $n>0$, $q\rightarrow -1$ when $a\rightarrow 0$ and $q\rightarrow -\infty$ when $a\rightarrow a_*$.

For $n<0$, $q\rightarrow -\infty$ when $a\rightarrow a_*$ and $q\rightarrow -1$ when $a\rightarrow +\infty$.

Some curves giving the evolution of  $q$  as a function of
the scale factor $a$ are plotted in Figs. \ref{newwNpos} and \ref{newwNneg}
for $n>0$ and $n<0$.

\subsection{The differential equation}
\label{sec_darkdiff}

The temporal  evolution of the scale factor $a(t)$ is determined by the Friedmann equation (\ref{basic2b}).
Introducing the normalized radius $R=a/a_*$, the density (\ref{d1}) can be written
\begin{equation}
\label{dark5}
\rho=\frac{\rho_*}{\lbrack 1-R^{3(1+\alpha)/n}\rbrack^n}.
\end{equation}
Substituting this expression in Eq. (\ref{basic2b}), we obtain the differential equation
\begin{equation}
\label{dark6}
\dot R=\frac{\epsilon K R}{\lbrack 1-R^{3(1+\alpha)/n}\rbrack^{n/2}},
\end{equation}
where $K=(8\pi G\rho_*/3)^{1/2}$ and $\epsilon=\pm 1$. In general, we shall select the sign $\epsilon=+1$ corresponding to an expanding universe ($\dot R>0$), except in the case of a bouncing universe where both signs of $\epsilon$ must be considered. The solution can be written as
\begin{equation}
\label{dark7}
\epsilon Kt=\int \left\lbrack 1-R^{3(1+\alpha)/n}\right\rbrack^{n/2}\, \frac{dR}{R},
\end{equation}
or, after a change of variables $x=R^{3(1+\alpha)/n}$, as
\begin{equation}
\label{dark8}
\frac{3(\alpha+1)}{n}\epsilon Kt=\int^{R^{3(\alpha+1)/n}} (1-x)^{n/2}\, \frac{dx}{x}.
\end{equation}
The integral can be expressed in terms of hypergeometric functions. Some simple analytical expressions can be obtained for specific values of $n$. Actually, we can have a good idea of the behavior of the solution of Eq. (\ref{dark6}) by considering asymptotic limits (see below). The complete solution is represented in the figures by solving Eq. (\ref{dark6}) numerically.

\subsection{The case $n>0$}
\label{sec_darkreppos}

The universe starts from $t\rightarrow -\infty$ with a vanishing radius $R=0$, a finite density $\rho=\rho_*$, and a finite pressure $p=-\rho_* c^2$. When $t\rightarrow -\infty$,
\begin{eqnarray}
\label{dark9}
R\sim Ae^{Kt}.
\end{eqnarray}
This corresponds to an exponential expansion (early inflation) due to the fact that the density is approximately constant. Then, the universe undergoes a finite time singularity at a time $t_{*}$. When $t\rightarrow t_{*}$, the radius tends to its maximum value $R=1$ while the density tends to $+\infty$ and the pressure to $-\infty$. This is a future singularity of type III. Close to the singularity, we have
\begin{eqnarray}
\label{dark10}
1-R\sim \left \lbrace \frac{2+n}{2}\left\lbrack \frac{n}{3(\alpha+1)}\right \rbrack^{n/2} K(t_{*}-t)\right \rbrace^{2/(2+n)},\qquad
\end{eqnarray}
\begin{eqnarray}
\label{marre1}
\frac{\rho}{\rho_*}\sim \left \lbrack \frac{3}{2}(1+\alpha)\frac{2+n}{n} K(t_{*}-t)\right \rbrack^{-2n/(2+n)}.
\end{eqnarray}

\begin{figure}[!h]
\begin{center}
\includegraphics[clip,scale=0.3]{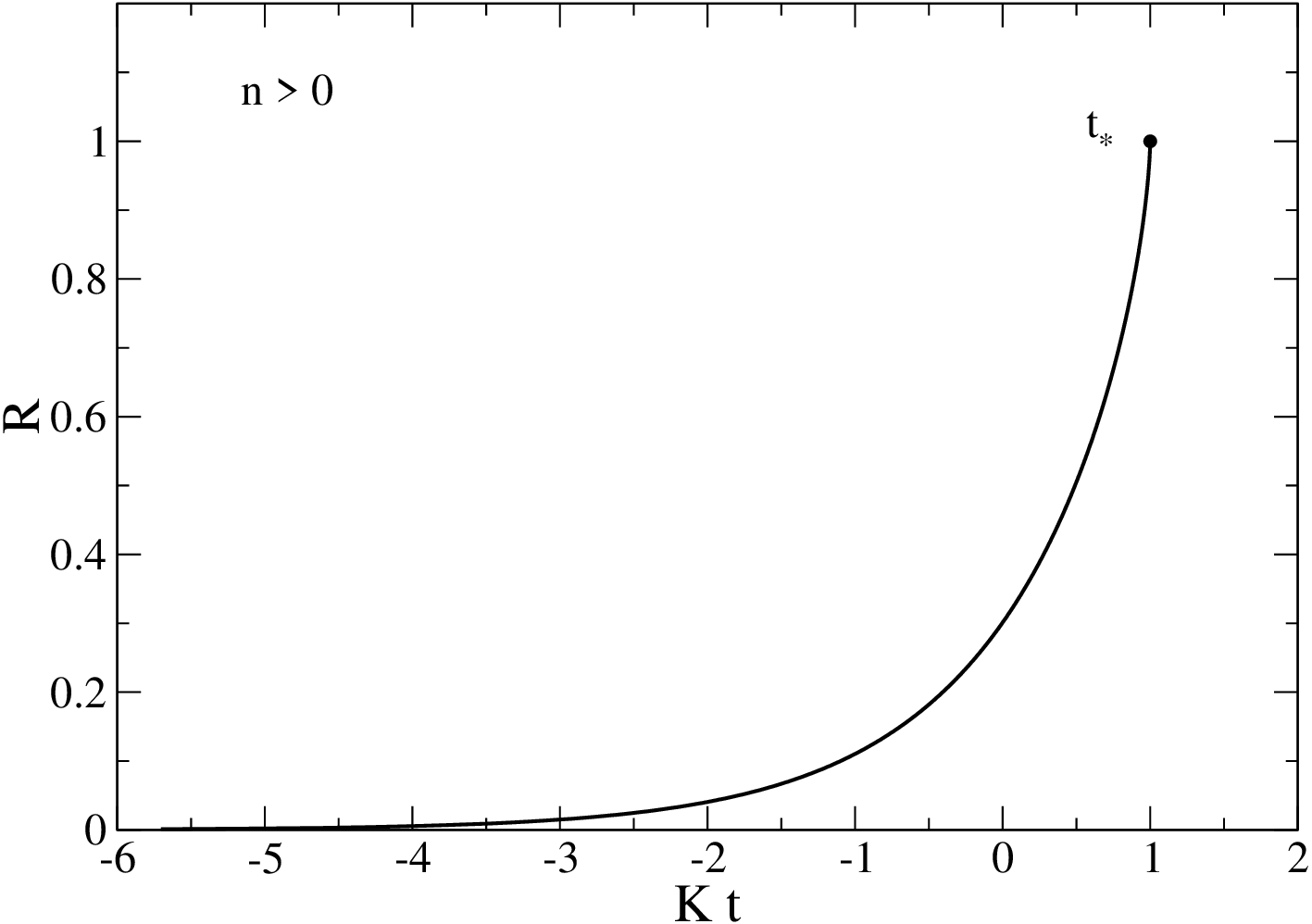}
\caption{Evolution of the radius $R$ as a function of time for $n>0$ (specifically $n=1$). We have taken $\alpha=0$. We have chosen the constant of integration such that $R=1$ at $t=t_{*}=1/K$.}
\label{newrayonNpos}
\end{center}
\end{figure}

The evolution of the scale factor is represented in Fig. \ref{newrayonNpos}. Some simple analytical results can be obtained in particular cases.

For $n=1$, using the identity
\begin{eqnarray}
\label{dark11}
\int \sqrt{1-x}\, \frac{dx}{x}=2\sqrt{1-x}+\ln\left (\frac{1-\sqrt{1-x}}{1+\sqrt{1-x}}\right ),
\end{eqnarray}
and for $n=2$, using the identity
\begin{eqnarray}
\label{dark12}
\int (1-x)\, \frac{dx}{x}=-x+\ln x,
\end{eqnarray}
we can obtain $t(R)$ from Eq. (\ref{dark8}).

\subsection{The case $n<0$}
\label{sec_darkrepneg}

The early evolution of the universe depends on the value of $n$. Different cases must be considered.

(i) For $n<-2$, the universe starts from $t\rightarrow -\infty$ with a finite radius $R=1$, a vanishing density $\rho=0$, and a vanishing pressure $p=0$ (past peculiarity). When $t\rightarrow -\infty$,
\begin{eqnarray}
\label{dark13}
R-1\sim \left \lbrace \frac{2}{|n|-2}\left\lbrack \frac{|n|}{3(\alpha+1)}\right \rbrack^{|n|/2} \frac{1}{(-Kt)}\right \rbrace^{2/(|n|-2)},\qquad
\end{eqnarray}
\begin{eqnarray}
\label{marre3}
\frac{\rho}{\rho_*}\sim \left \lbrack \frac{2}{3}\frac{1}{\alpha+1}\frac{|n|}{|n|-2}\frac{1}{(-Kt)}\right \rbrack^{2|n|/(|n|-2)},
\end{eqnarray}
The density tends to zero algebraically rapidly.

(ii) For $n=-2$, the universe starts from $t\rightarrow -\infty$ with a finite radius $R=1$, a vanishing density $\rho=0$, and a vanishing pressure $p=0$ (past peculiarity). When $t\rightarrow -\infty$,
\begin{eqnarray}
\label{dark14}
R-1\sim Ae^{\frac{3(1+\alpha)}{2}Kt},
\end{eqnarray}
\begin{eqnarray}
\label{marre4}
\frac{\rho}{\rho_*}\sim \frac{9}{4}(1+\alpha)^2A^2e^{{3(1+\alpha)}Kt}.
\end{eqnarray}
The density tends to zero exponentially rapidly.

(iii) For $-2<n<0$, the universe starts at $t=0$ with a finite radius $R=1$ and a vanishing density $\rho=0$ (past peculiarity). When $t\rightarrow 0$,
\begin{eqnarray}
\label{dark15}
R-1\sim \left \lbrace \frac{2-|n|}{2}\left\lbrack \frac{3(\alpha+1)}{|n|}\right \rbrack^{|n|/2} K t\right \rbrace^{2/(2-|n|)},
\end{eqnarray}
\begin{eqnarray}
\label{marre2}
\frac{\rho}{\rho_*}\sim \left \lbrack \frac{3}{2}(1+\alpha) \frac{2-|n|}{|n|} K t\right \rbrack^{2|n|/(2-|n|)}.
\end{eqnarray}
At $t=0$, the pressure vanishes for $-2<n<-1$, is finite for $n=-1$ and tends to $-\infty$ for $n>-1$. In this last case, there is a past singularity of type II. Actually, we can extend the solution to $t<0$ (except, maybe, in the case $n>-1$ where the pressure diverges). This describes a phase of contraction of the universe, corresponding to the solution of Eq. (\ref{dark6}) with $\epsilon=-1$. This leads to a model of bouncing phantom universe that collapses for $t<0$ (with decreasing density), disappears at $t=0$ (the density vanishes), and expands for $t>0$ (with increasing density).

On the other hand, for $t\rightarrow +\infty$, the density tends to a constant $\rho_*$, implying an exponential growth of the scale factor as
\begin{eqnarray}
\label{dark16}
R\sim A' e^{Kt}.
\end{eqnarray}
This corresponds to a phase of late inflation. The pressure $p\rightarrow -\rho_* c^2$.

\begin{figure}[!h]
\begin{center}
\includegraphics[clip,scale=0.3]{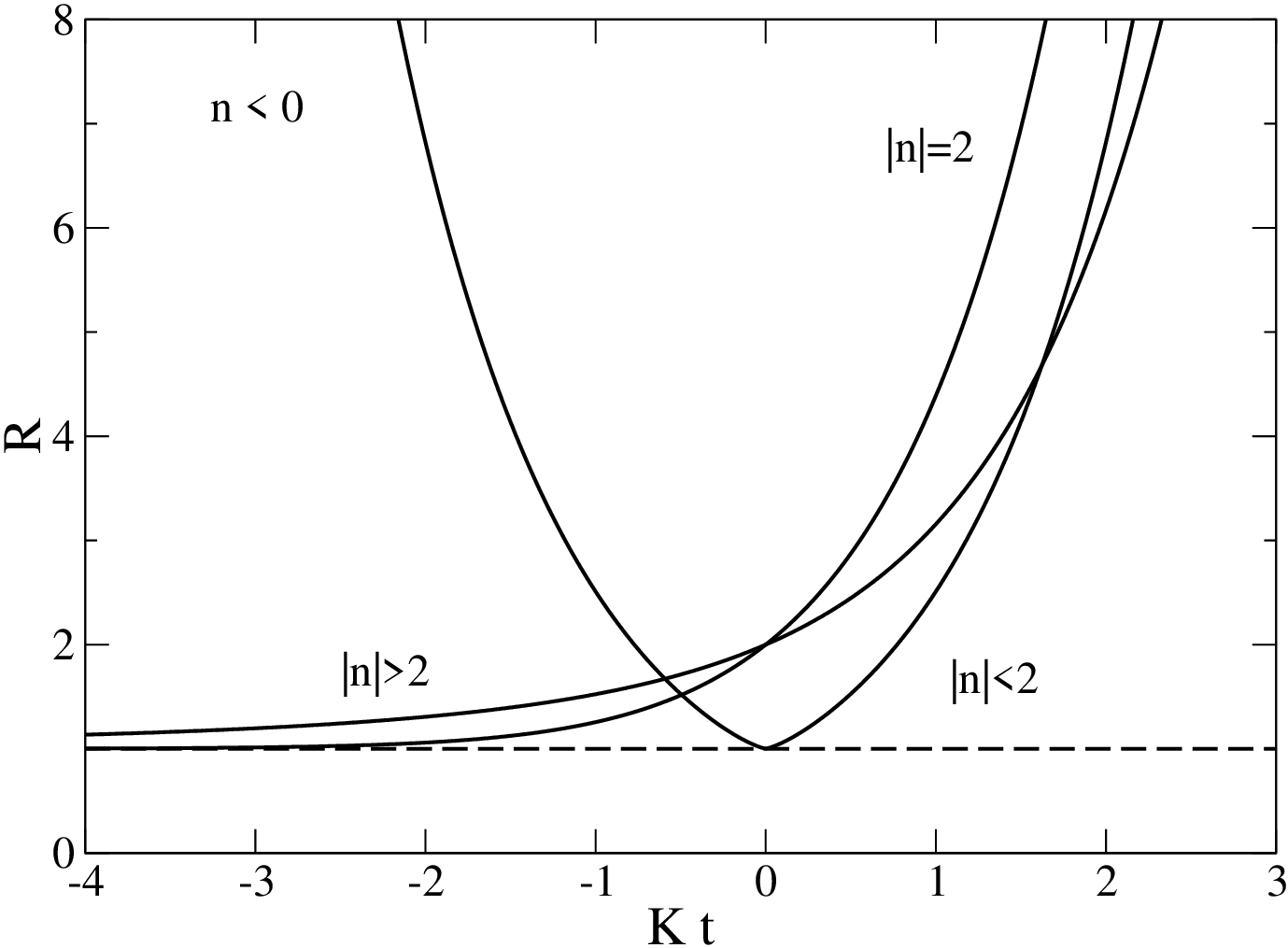}
\caption{Evolution of the radius $R$ as a function of time for $n<-2$, $n=-2$ and $-2<n<0$ (specifically $n=-3$, $n=-2$ and $n=-1/2$). We have taken $\alpha=0$. For $n\le -2$, we have chosen the constant of integration such that $R=2$ at $t=0$. For $-2<n<0$, the constant of integration has been chosen such that $R=1$ at $t=0$. In that case, we have a bouncing universe (see text for details). }
\label{newrayonNneg}
\end{center}
\end{figure}

The evolution of the scale factor is represented in Fig. \ref{newrayonNneg} for $n<0$. Some simple analytical results can be obtained in particular cases.

\begin{figure}[!h]
\begin{center}
\includegraphics[clip,scale=0.3]{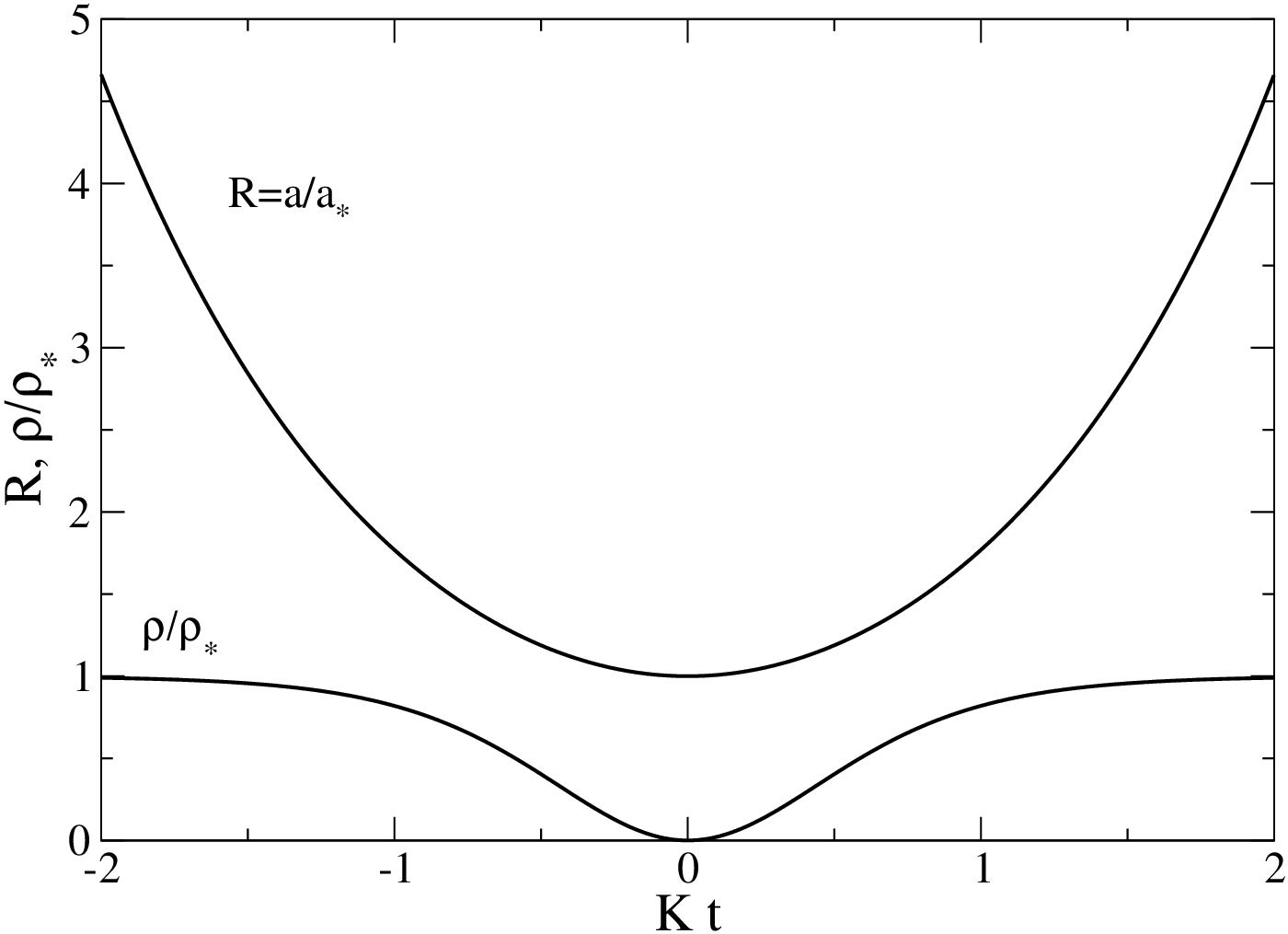}
\caption{Analytical model of a bouncing phantom universe corresponding to $n=-1$. We have taken $\alpha=0$. The universe starts from $t=-\infty$ with a maximum  density $\rho_*$. It first experiences a phase of contraction during which its density decreases. At $t=0$, it reaches its minimum radius $R=1$ and its density vanishes (the universe ``disappears''). For $t>0$, the universe expands and its density increases up to the maximum value $\rho_*$.  }
\label{bounce}
\end{center}
\end{figure}

For $n=-1$, using the identity
\begin{eqnarray}
\label{dark17}
\int \frac{1}{\sqrt{1-x}}\, \frac{dx}{x}=\ln\left (\frac{1-\sqrt{1-x}}{1+\sqrt{1-x}}\right ),
\end{eqnarray}
we obtain
\begin{eqnarray}
\label{dark18}
R=\cosh^{2/\lbrack 3(1+\alpha)\rbrack}\left\lbrack \frac{3}{2}(1+\alpha)Kt\right\rbrack,
\end{eqnarray}
\begin{eqnarray}
\label{tata1}
\frac{\rho}{\rho_*}=\tanh^2\left\lbrack \frac{3}{2}(1+\alpha)Kt\right\rbrack.
\end{eqnarray}
This provides an analytical solution of a bouncing phantom universe (see Fig. \ref{bounce}). We can explicitly check that Eq. (\ref{dark18}) has the asymptotic forms (\ref{dark15}) and (\ref{dark16}) with $A'=2^{-2/\lbrack 3(1+\alpha)\rbrack}$. For $\alpha=0$, this model has a constant negative pressure $p=-|k|c^2$. It belongs therefore to the same ``class'' as the $\Lambda$CDM model (Paper I) and the anti-$\Lambda$CDM model (Paper II) that also have a constant pressure. These three models admit simple analytical expressions.

For $n=-2$, using the identity
\begin{eqnarray}
\label{dark19}
\int \frac{1}{1-x}\, \frac{dx}{x}=\ln \left (\frac{x}{1-x}\right ),
\end{eqnarray}
we obtain
\begin{eqnarray}
\label{dark20}
R=\left\lbrack 1+e^{\frac{3}{2}(1+\alpha)K t}\right\rbrack^{2/\lbrack 3(1+\alpha)\rbrack},
\end{eqnarray}
\begin{eqnarray}
\label{tata2}
\frac{\rho}{\rho_*}=\frac{1}{\left\lbrack 1+e^{-\frac{3}{2}(1+\alpha)Kt}\right \rbrack^2}.
\end{eqnarray}
This provides an analytical solution of a phantom universe exhibiting a past peculiarity and a late inflation. Equation (\ref{dark20}) has the asymptotic forms (\ref{dark14}) and (\ref{dark16}) with $A'=1$ and $A=2/\lbrack 3(1+\alpha)\rbrack$.

For $n=-1/2$, we have  the identity
\begin{eqnarray}
\label{tutu2}
\int \frac{1}{(1-x)^{1/4}}\, \frac{dx}{x}=2 \arctan\left\lbrack (1-x)^{1/4}\right\rbrack\nonumber\\
+\ln\left\lbrack \frac{1-(1-x)^{1/4}}{1+(1-x)^{1/4}}\right\rbrack,
\end{eqnarray}
which determines $t(R)$ using Eq. (\ref{dark8}). For $\alpha=0$, this solution corresponds to the phantom Chaplygin gas.

{\it Remark:} We note that the solution with $n\le -2$ looks similar to the Eddington-Lema\^itre model (see Fig. 1 in Paper I) since the universe is ``static'' in the past with a finite radius $R=1$ and expands exponentially rapidly in the future. However, in the Eddington-Lema\^itre model, the density decreases with time while, in the present (phantom) model, it increases with time. In addition, in the Eddington-Lema\^itre model, the density tends to a finite value when $t\rightarrow -\infty$ while in the present model, it tends to zero. Therefore, these models are physically very different.

\section{Scalar field models}
\label{sec_scalar}

In this section, we introduce a representation of the phantom universe in terms of scalar field models. We determine the potential of the scalar field corresponding to the equation of state (\ref{basic3}) using the general methodology exposed in \cite{cst}. We consider a normal scalar field  and a tachyon field. Although totally equivalent to fluid equations, this scalar field representation may be useful in order to make the link with more fundamental theories, like those arising in particle physics and string theory \cite{saridakis}.

\subsection{Phantom scalar field}
\label{sec_phantom}

A fluid with an equation of state parameter satisfying $w>-1$ can be described in terms of an ordinary scalar field minimally coupled to gravity called a quintessence field \cite{quintessence}. A fluid with an equation of state parameter satisfying $w<-1$ can be described in terms of a phantom scalar field. The phantom field can be obtained from the quintessence field by making the transformation $\phi\rightarrow i\phi$. As a result, the phantom scalar field evolves according to the equation
\begin{equation}
\label{quintessence1}
\ddot \phi+3H\dot\phi-\frac{dV}{d\phi}=0,
\end{equation}
where $V(\phi)$ is the potential of the scalar field. The density and the pressure are given by
\begin{equation}
\label{quintessence2}
\rho c^2=-\frac{1}{2}\dot\phi^2+V(\phi),\qquad p=-\frac{1}{2}\dot\phi^2-V(\phi).
\end{equation}
We note that the sign of the kinetic term is reversed with respect to the quintessence scalar field. This implies that the phantom scalar field tends to run up, not down, the potential towards larger energies.

From Eq. (\ref{quintessence2}), we get
\begin{equation}
\label{quintessence3}
\dot\phi^2=|1+w|\rho c^2,
\end{equation}
where we have written $p=w\rho c^2$. Using $\dot\phi=(d\phi/da) H a$, and the Friedmann equation (\ref{basic2b}) valid for a flat universe, we obtain
\begin{equation}
\label{quintessence4}
\frac{d\phi}{da}=\left (\frac{3c^2}{8\pi G}\right )^{1/2}\frac{\sqrt{|1+w|}}{a}.
\end{equation}
For the equation of state (\ref{basic3}), using Eqs. (\ref{d1}) and (\ref{w2}), and setting $R=a/a_*$, we can rewrite Eq. (\ref{quintessence4}) in the form
\begin{equation}
\label{quintessence5}
\frac{d\phi}{dR}=\left (\frac{3c^2}{8\pi G}\right )^{1/2}\frac{\sqrt{\alpha+1}}{R}\frac{R^{3(1+\alpha)/2n}}{\sqrt{1-R^{3(1+\alpha)/n}}}.
\end{equation}
With the change of variables
\begin{equation}
\label{quintessence6}
x=R^{3(\alpha+1)/2n},\qquad \psi=\left (\frac{8\pi G}{3c^2}\right )^{1/2}\frac{3\sqrt{\alpha+1}}{2n}\phi,
\end{equation}
we find that
\begin{equation}
\label{quintessence7}
\psi=\int \frac{dx}{\sqrt{1-x^2}}={\rm Arcsin} (x).
\end{equation}
On the other hand, according to Eq. (\ref{quintessence2}), we have
\begin{equation}
\label{quintessence10}
V=\frac{1}{2}(1-w)\rho c^2.
\end{equation}
For the equation of state (\ref{basic3}), using Eqs. (\ref{d1}) and (\ref{w2}), we obtain
\begin{equation}
\label{quintessence11}
V=\frac{1}{2}\rho_* c^2 \frac{2-(1-\alpha)x^2}{(1-x^2)^{n+1}}.
\end{equation}
Since $x=\sin\psi$, the scalar field potential is explicitly given by
\begin{equation}
\label{quintessence14}
V(\psi)=\frac{1}{2}\rho_* c^2 \frac{(1-\alpha)\cos^2\psi+\alpha+1}{\cos^{2(n+1)}\psi},
\end{equation}
and $R^{3(\alpha+1)/2n}=\sin\psi$. In these models, $0\le \psi\le \pi/2$.

The case $\alpha=-1$ and $k<0$ (see Appendix \ref{sec_eosgm}) must be treated specifically. Repeating the preceding procedure, we find that the potential of the scalar field is
\begin{eqnarray}
\label{quintessence22}
V(\phi)=\rho_* c^2 \left (\frac{|n| c^2}{2\pi G}\right )^n \left (1+\frac{n^2c^2}{12\pi G\phi^2}\right )\frac{1}{\phi^{2n}},
\end{eqnarray}
\begin{equation}
\label{quintessence23}
-\ln R=\frac{2\pi G}{nc^2}\phi^2.
\end{equation}
In these models $\phi\ge 0$.

Finally, for a linear equation of state $p=\alpha\rho c^2$ with $\alpha<-1$, writing the relation between the density and the scale factor as  $\rho/\rho_*=(a/a_*)^{3|1+\alpha|}$, we obtain \cite{cst}:
\begin{equation}
\label{quintessence24}
V(\phi)=\frac{1}{2}\rho_* c^2 (1-\alpha) e^{3\sqrt{|\alpha+1|}\left (\frac{8\pi G}{3c^2}\right )^{1/2}\phi},
\end{equation}
\begin{equation}
\label{quintessence25}
\phi=\left (\frac{3c^2}{8\pi G}\right )^{1/2}\sqrt{|1+\alpha|}\, \ln R,
\end{equation}
where $R=a/a_*$ and $\phi\le 0$. Since $R\propto (t_s-t)^{-2/(3|1+\alpha|)}$, the scalar field evolves in time as $\phi=-(c^2/6\pi G |1+\alpha|)^{1/2}\ln (t_s-t)$.

\subsection{Phantom tachyon field}
\label{sec_tachyon}

Performing the transformation $\phi\rightarrow i\phi$ in the equations of an ordinary tachyon field \cite{tachyon}, we find that a phantom tachyon field evolves according to the equation
\begin{equation}
\label{tachyon1}
\frac{\ddot \phi}{1+\dot\phi^2}+3H\dot\phi-\frac{1}{V}\frac{dV}{d\phi}=0.
\end{equation}
The density and the pressure are given by
\begin{equation}
\label{tachyon2}
\rho c^2=\frac{V(\phi)}{\sqrt{1+\dot\phi^2}},\qquad p=-V(\phi)\sqrt{1+\dot\phi^2}.
\end{equation}
From these equations, we obtain
\begin{equation}
\label{tachyon3}
\dot\phi^2=|1+w|,
\end{equation}
where we have written $p=w\rho c^2$. Using $\dot\phi=(d\phi/da) H a$, and the Friedmann equation (\ref{basic2b}), we get
\begin{equation}
\label{tachyon4}
\frac{d\phi}{da}=\left (\frac{3c^2}{8\pi G}\right )^{1/2}\frac{\sqrt{|1+w|}}{\sqrt{\rho c^2} a}.
\end{equation}
For the equation of state (\ref{basic3}), using Eqs. (\ref{d1}) and (\ref{w2}), we can rewrite Eq. (\ref{tachyon4}) in the form
\begin{eqnarray}
\label{tachyon5}
\frac{d\phi}{dR}=\frac{1}{\sqrt{\rho_*c^2}}\left (\frac{3c^2}{8\pi G}\right )^{1/2}\frac{\sqrt{\alpha+1}}{R}R^{3(1+\alpha)/2n}\nonumber\\
\times\left\lbrack 1-R^{3(1+\alpha)/n}\right\rbrack^{(n-1)/2}.
\end{eqnarray}
With the change of variables
\begin{equation}
\label{tachyon6}
x=R^{3(\alpha+1)/2n},\qquad \psi=\sqrt{\rho_*c^2}\left (\frac{8\pi G}{3c^2}\right )^{1/2}\frac{3\sqrt{1+\alpha}}{2n}\phi,
\end{equation}
we find that
\begin{equation}
\label{tachyon7}
\psi=\int (1-x^2)^{(n-1)/2}\, {dx}.
\end{equation}
On the other hand, from Eq. (\ref{tachyon2}), we have
\begin{equation}
\label{tachyon8}
V^2=-w\rho^2 c^4.
\end{equation}
For the equation of state (\ref{basic3}), using Eqs. (\ref{d1}) and (\ref{w2}), we obtain
\begin{equation}
\label{tachyon9}
V^2=\rho_*^2 c^4 \frac{\alpha x^2+1}{(1-x^2)^{2n+1}}.
\end{equation}
Therefore, the scalar field potential $V(\psi)$ is given in parametric form by Eqs. (\ref{tachyon7}) and (\ref{tachyon9}). Let us consider particular cases.

(i) For $n=1$, we find that $x=\psi$. Therefore, we obtain
\begin{equation}
\label{tachyon10}
V^2=\rho_*^2 c^4 \frac{\alpha \psi^2+1}{(1-\psi^2)^{3}},
\end{equation}
and $R^{3(\alpha+1)/2}=\psi$ with $0\le \psi\le 1$.

(ii) For $n=-1$, we find that $x=\tanh\psi$. Therefore, we obtain
\begin{equation}
\label{tachyon15}
V^2=\rho_*^2 c^4 \frac{\alpha \tanh^2\psi+1}{\cosh^2\psi},
\end{equation}
and $R^{3(\alpha+1)/2}=1/\tanh\psi$ with $\psi\ge 0$.

(iii) For $n=-2$, we find that $x=\psi/\sqrt{1+\psi^2}$. Therefore, we obtain
\begin{equation}
\label{tachyon22}
V^2=\rho_*^2 c^4 \frac{(\alpha+1)\psi^2+1}{(1+\psi^2)^4},
\end{equation}
and $R^{3(\alpha+1)/4}=\sqrt{1+\psi^2}/\psi$
with $\psi\ge 0$.

(iv) For $n=-1/2$ and $\alpha=0$ (phantom Chaplygin gas), we find that $V(\phi)=\rho_* c^2$ is constant.

The case $\alpha=-1$ and $k>0$ (see Appendix \ref{sec_eosgm}) must be treated specifically. Repeating the preceding procedure, we find that the potential of the scalar field is
\begin{eqnarray}
\label{tachyon29}
V(\phi)^2=\rho_*^2 c^4 \left \lbrack \frac{|n|}{2\pi G\rho_* (n+1)^2}\right \rbrack^{2n/(n+1)}\frac{1}{\phi^{4n/(n+1)}}\nonumber\\
\times\left \lbrace 1+\frac{|n|}{3}\left \lbrack \frac{|n|}{2\pi G\rho_* (n+1)^2}\right \rbrack^{1/(n+1)}\frac{1}{\phi^{2/(n+1)}}\right \rbrace,\nonumber\\
\end{eqnarray}
\begin{equation}
\label{tachyon30}
-\ln R={\rm sgn}(n) \left \lbrack \frac{2\pi G \rho_* (n+1)^2}{|n|}\right \rbrack^{1/(n+1)}\phi^{2/(n+1)}.
\end{equation}
In these models $\phi\ge 0$.

Finally, for a linear equation of state $p=\alpha\rho c^2$ with $\alpha<-1$, writing the relation between the density and the scale factor as  $\rho/\rho_*=(a/a_*)^{3 |1+\alpha|}$, we obtain \cite{cst}:
\begin{equation}
\label{tachyon27}
V(\phi)=\frac{\sqrt{-\alpha}}{|1+\alpha|}\frac{c^2}{6\pi G}\frac{1}{\phi^2},
\end{equation}
\begin{equation}
\label{tachyon28}
\phi=-\frac{2}{3}\frac{1}{\sqrt{\rho_* c^2}} \left (\frac{3c^2}{8\pi G}\right )^{1/2}\frac{1}{\sqrt{|1+\alpha|}}R^{-3|1+\alpha|/2},
\end{equation}
where $R=a/a_*$ and $\phi\le 0$. Since $\rho=\rho_* R^{3|1+\alpha|}=1/\lbrack 6\pi G(1+\alpha)^2\rbrack (t_s-t)^{-2}$, the scalar field evolves in time as $\phi=-\sqrt{|1+\alpha|} \, (t_s-t)$.

\section{Conclusion}

In this paper, we have performed an exhaustive study of the
generalized equation of state (\ref{basic3}) in the case where the
pressure increases with the scale factor. This corresponds to the
so-called phantom cosmology \cite{caldwell}. 

The case $\alpha=-1$ was previously treated in 
\cite{stefancic}. For $n<-2$, the universe experiences a 
future singularity of type I (Big Rip): The
scale factor and the density diverge at a finite time. For $-2\le
n<0$, the scale factor and the density diverge in infinite time
(Little Rip). For $n>0$, the universe experiences a future singularity
of type III: The density diverges at a finite time while the scale
factor tends to a constant.

We have found that when $\alpha>-1$, the universe 
does not experience a Big Rip singularity. For $n<0$, there is a phase
of late inflation and, for $n>0$, the universe experiences a future
singularity of type III. The past evolution of the universe is
interesting. For $n>0$, there is a phase of early inflation. For $n<0$,
the universe exhibits a past peculiarity since the density vanishes
while the scale factor tends to a finite value. For $n\le -2$, the
evolution of the scale factor is similar to the Eddington-Lema\^itre
model (the universe is static in the infinite past and grows
exponentially rapidly in the future) but the evolution of the density
is very different (it starts from zero in the infinite past and
increases as the universe expands). For $-2<n<0$, we have obtained a
model of bouncing universe which also possesses peculiar features (the
density decreases in the past, vanishes at $t=0$, and increases in the
future).  For $-1<n<0$, this bouncing universe presents a past
singularity of type II since the pressure at $t=0$ is infinite while
the scale factor is finite and the density vanishes. For $n=-1$,
corresponding to a constant negative pressure, the bouncing phantom
universe admits a simple analytical expression.

Of course, most of these models are academic, and do not correspond to
the true evolution of our universe. However, we believe that it is
important to study the equation of state (\ref{basic3}) in full
generality. On the other hand, there are indications
\cite{observations} that the equation of state parameter $w$ of our
universe may become less than $-1$ in the close future (or even at
present), so the {\it late} evolution of the phantom models described
in this paper may be physically relevant.

A drawback of the simple form of phantom cosmology considered in this paper is that it does not connect smoothly to the matter era (which has $w=0$). Therefore, we cannot realistically extend the phantom models to the past and obtain unified models of dust matter and phantom dark energy ($w<-1$), contrary to the unified models of dust matter and quintessence dark energy ($w>-1$) based on the generalized Chaplygin gas considered in Paper II.  A unification of dust matter and phantom dark energy can be achieved in more general models allowing to cross the phantom divide \cite{divide}. This generalization assumes an interaction between dark matter and dark energy. These models are very interesting because they may provide a solution to the ``cosmic coincidence problem'' (the fact that the ratio of dark matter and dark energy is of order one).

The phantom cosmology is also interesting for its connection to Hoyle's version of the steady state theory \cite{hoyle}, for its connection to wormholes \cite{wormholes}, and for its very strange thermodynamics allowing for the existence of negative temperatures \cite{thermophantom} like in 2D turbulence \cite{onsager}.

However, we may recall that there is no firm evidence that we live in a phantom universe. The model of Paper II, corresponding to the standard $\Lambda$CDM model with the primordial singularity removed, may correctly describe the whole evolution of our universe. Therefore, a more precise determination of the equation of state parameter $w$ will help discriminate between these different models.

\appendix

\section{Equation of state $p=(-\rho+k\rho^{\gamma})c^2$ with $k<0$}
\label{sec_eosgm}

In this Appendix, we specifically study the equation of state (\ref{basic3}) with $\alpha=-1$ and $k<0$, namely
\begin{eqnarray}
\label{eosgm0}
p=(-\rho-|k|\rho^\gamma)c^2.
\end{eqnarray}
Since $w<-1$, this equation of state describes a phantom universe. This equation of state was introduced by  Nojiri \& Odintsov \cite{quantum} and studied by Stefanci\'c \cite{stefancic}. Nojiri {\it et al.} \cite{classification} used it to illustrate their classification  of future finite time singularities. For the completeness of our study, we shall re-derive their results in a more compact form (with our notations) and give a few complements. We follow the same presentation as in Papers I and II.

\subsection{The case $n>0$}

The equation of continuity  (\ref{basic1}) can be integrated into
\begin{eqnarray}
\label{eosgm1}
\rho=\frac{\rho_*}{\ln(a_*/a)^n},
\end{eqnarray}
where $\rho_*=(n/3|k|)^n$ and $a_*$ is a constant of integration.  The density is defined for $a\le a_*$. When $a\rightarrow 0$, $\rho\rightarrow 0$ and $p\rightarrow 0$; when $a\rightarrow a_*$, $\rho\rightarrow +\infty$ and $p\rightarrow -\infty$.

\begin{figure}[!h]
\begin{center}
\includegraphics[clip,scale=0.3]{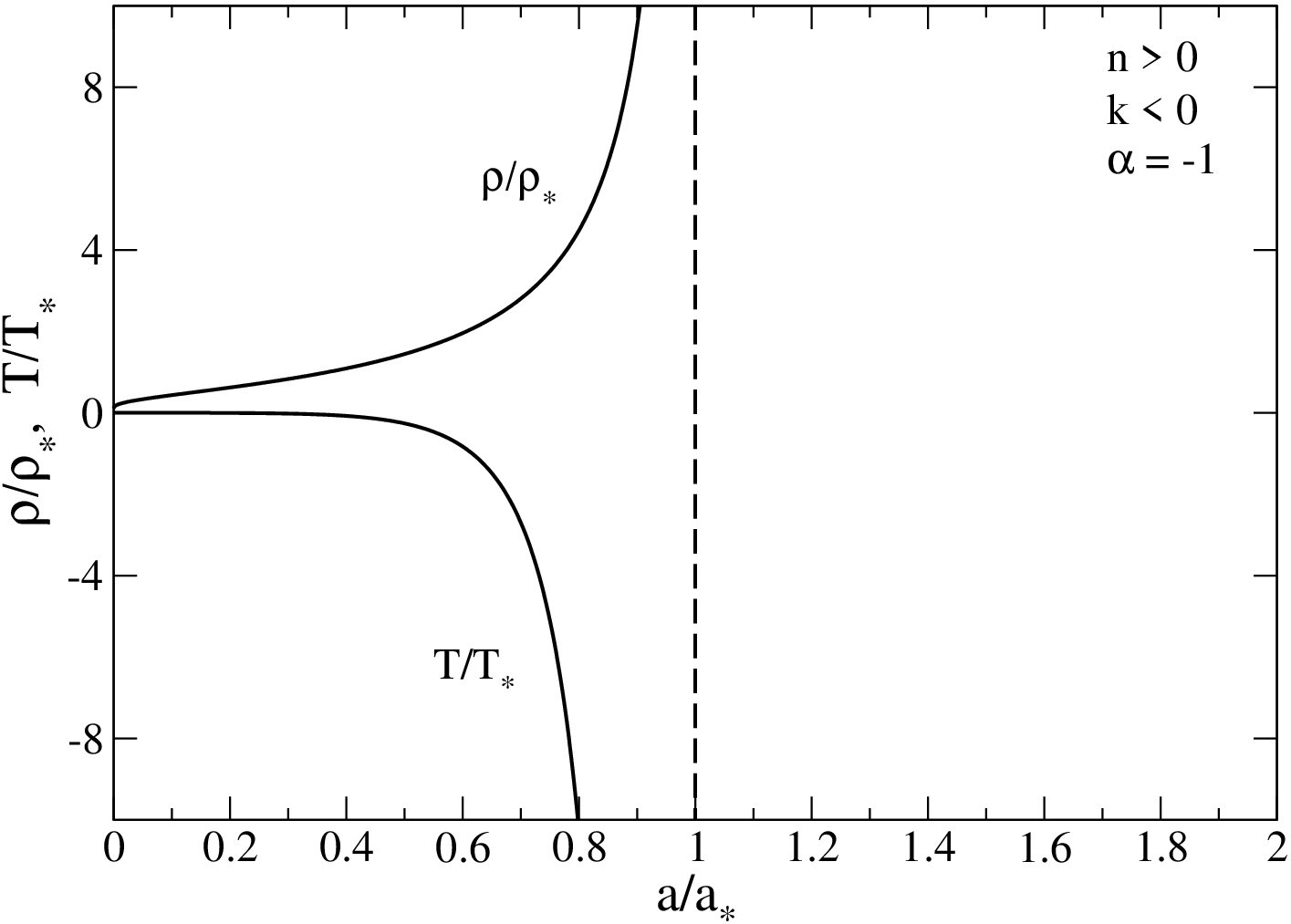}
\caption{Evolution of the density and temperature as a function of the scale factor. We have taken $n=1$.}
\label{newannexedensityPOS}
\end{center}
\end{figure}

The thermodynamical equation (\ref{a3}) can be integrated into
\begin{eqnarray}
\label{eosgm2}
T=-T_* \left (\frac{\rho}{\rho_*}\right )^{(n+1)/n} e^{-3(\rho_*/\rho)^{1/n}},
\end{eqnarray}
where $T_*>0$ is a constant of integration. Combined with Eq. (\ref{eosgm1}), we obtain
\begin{eqnarray}
\label{eosgm3}
T=- \frac{T_*}{\ln(a_*/a)^{n+1}}\left (\frac{a}{a_*}\right )^3.
\end{eqnarray}
When $a\rightarrow a_*$, $T\rightarrow 0$; when $a\rightarrow a_*$, $T\rightarrow -\infty$.
The evolution of the density and temperature as a function of the scale factor is represented in Fig. \ref{newannexedensityPOS}.

The equation of state can be written as $p=w\rho c^2$ with
\begin{equation}
\label{eosgm5}
w=-1-\frac{n}{3}\left (\frac{\rho}{\rho_*}\right )^{1/n}.
\end{equation}
When $a\rightarrow 0$, $w\rightarrow -1$; when  $a\rightarrow a_*$, $w\rightarrow -\infty$.

The deceleration parameter is given by Eqs. (I-77) and (I-78). Together with Eq. (\ref{eosgm5}), we obtain
\begin{equation}
\label{eosgm7}
q=-1-\frac{n}{2}\left (\frac{\rho}{\rho_*}\right )^{1/n}.
\end{equation}
When $a\rightarrow 0$,  $q\rightarrow -1$; when $a\rightarrow a_*$, $q\rightarrow -\infty$.

The velocity of sound is given by
\begin{equation}
\label{eosgm9}
\frac{c_s^2}{c^2}=-1-\frac{n+1}{3}\left (\frac{\rho}{\rho_*}\right )^{1/n}.
\end{equation}
When $a\rightarrow 0$, $(c_s/c)^2\rightarrow -1$; when $a\rightarrow a_*$, $(c_s/c)^2\rightarrow -\infty$. The velocity of sound is always imaginary.

\begin{figure}[!h]
\begin{center}
\includegraphics[clip,scale=0.3]{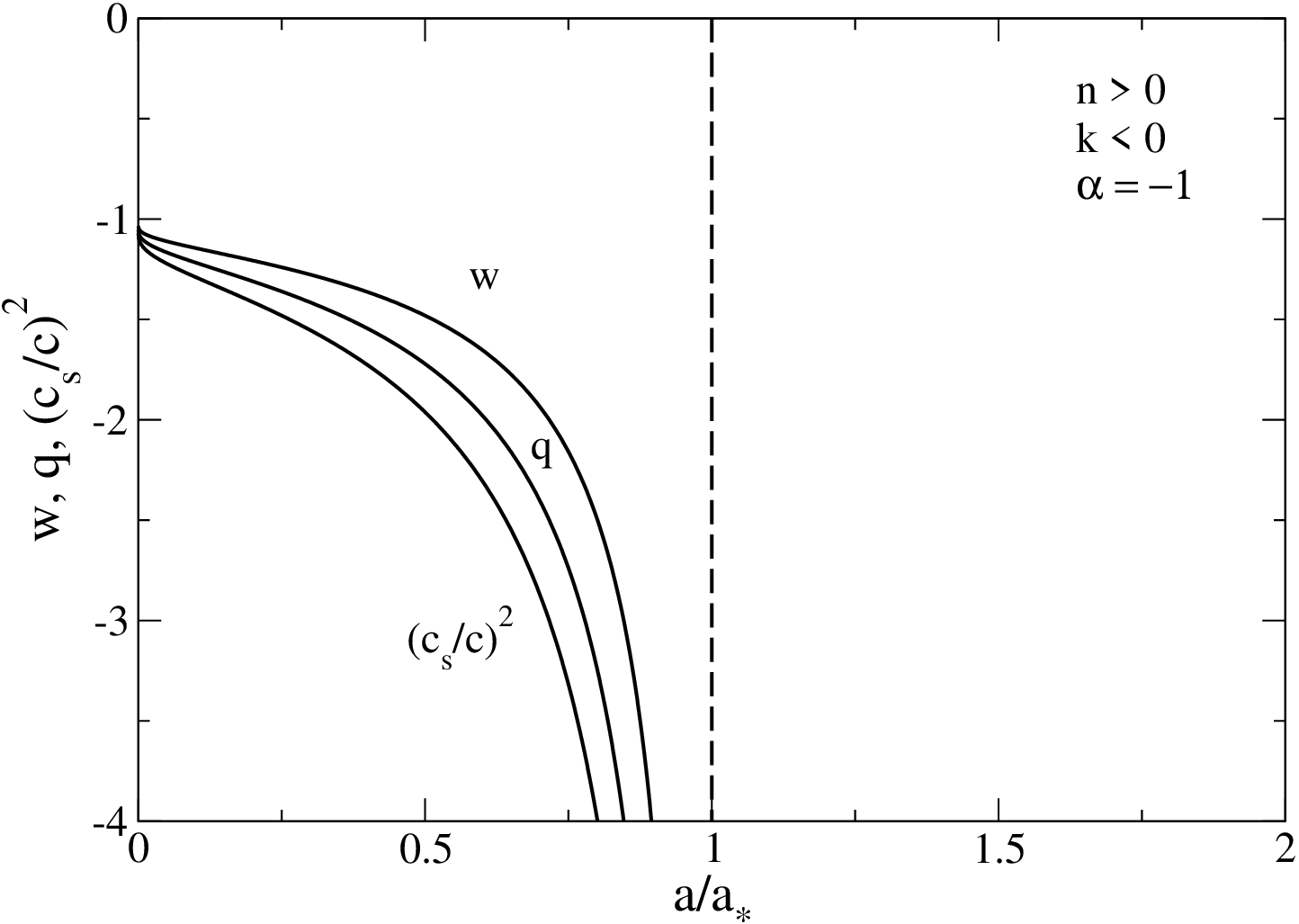}
\caption{Evolution of $w$, $q$, and $(c_s/c)^2$ as a function of the scale factor $a$. We have taken $n=1$. }
\label{newannexewPOS}
\end{center}
\end{figure}

The evolution of $w$, $q$, and $(c_s/c)^2$ as a function of the scale factor $a$ is represented in Fig. \ref{newannexewPOS}.

Setting $R=a/a_*$, the Friedmann equation (\ref{basic2b}) can be written
\begin{eqnarray}
\label{eosgm12}
\dot R=\frac{KR}{(-\ln R)^{n/2}},
\end{eqnarray}
where $K=(8\pi G\rho_*/3)^{1/2}$. Its solution is
\begin{eqnarray}
\label{eosgm13}
R(t)=e^{-\left \lbrack \frac{2+n}{2} K(t_{*}-t)\right \rbrack^{2/(2+n)}},
\end{eqnarray}
\begin{eqnarray}
\label{marre5}
\frac{\rho(t)}{\rho_*}=\left \lbrack \frac{2+n}{2} K(t_{*}-t)\right \rbrack^{-2n/(2+n)}.
\end{eqnarray}
The universe starts from $t=-\infty$ with a vanishing radius $R=0$, a vanishing density $\rho=0$ and a vanishing pressure $p=0$. This corresponds to a generalized past peculiarity. It also undergoes a future singularity of type III: At $t=t_{*}$, the density tends to $+\infty$ and the pressure tends to $-\infty$ while the radius reaches its maximum value $R=1$.

\begin{figure}[!h]
\begin{center}
\includegraphics[clip,scale=0.3]{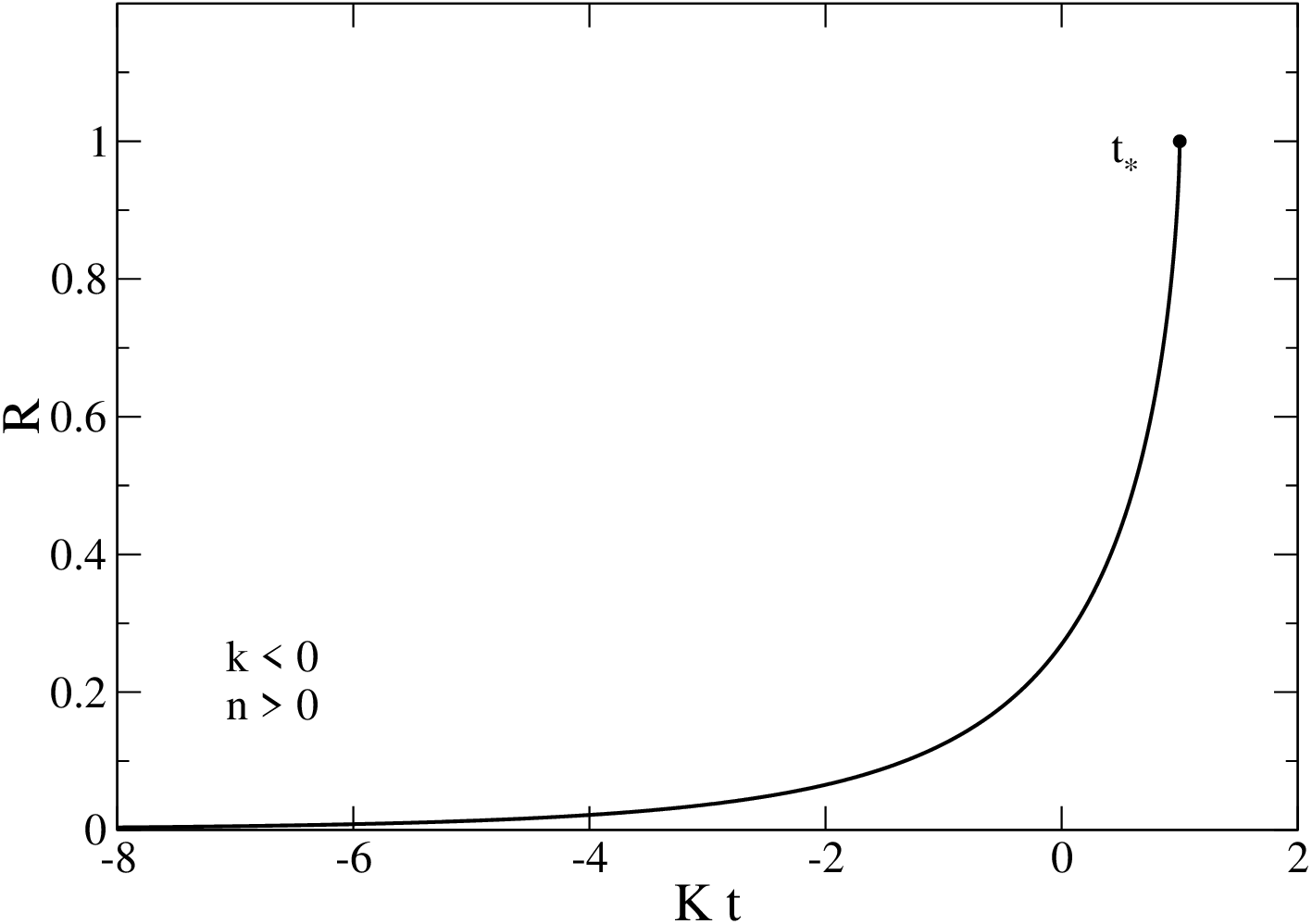}
\caption{Evolution of the radius with time for $n>0$ (specifically $n=1$). We have arbitrarily taken $t_{*}=1$.}
\label{annexeKnegNpos}
\end{center}
\end{figure}

The evolution of the scale factor is represented in Fig. \ref{annexeKnegNpos}.

\subsection{The case $n<0$}

The equation of continuity  (\ref{basic1}) can be integrated into
\begin{eqnarray}
\label{eosgm14}
\rho=\frac{\rho_*}{\ln(a/a_*)^n},
\end{eqnarray}
where $\rho_*=(|n|/3|k|)^n$ and $a_*$ is a constant of integration.  The density is defined for $a\ge a_*$. When $a\rightarrow a_*$, $\rho\rightarrow 0$. In the same limit, $p\rightarrow 0$ for $n<-1$, $p$ tends to a finite value for $n=-1$, and  $p \rightarrow -\infty$ for $n>-1$.  When $a\rightarrow +\infty$, $\rho\rightarrow +\infty$, and $p\rightarrow -\infty$. 

\begin{figure}[!h]
\begin{center}
\includegraphics[clip,scale=0.3]{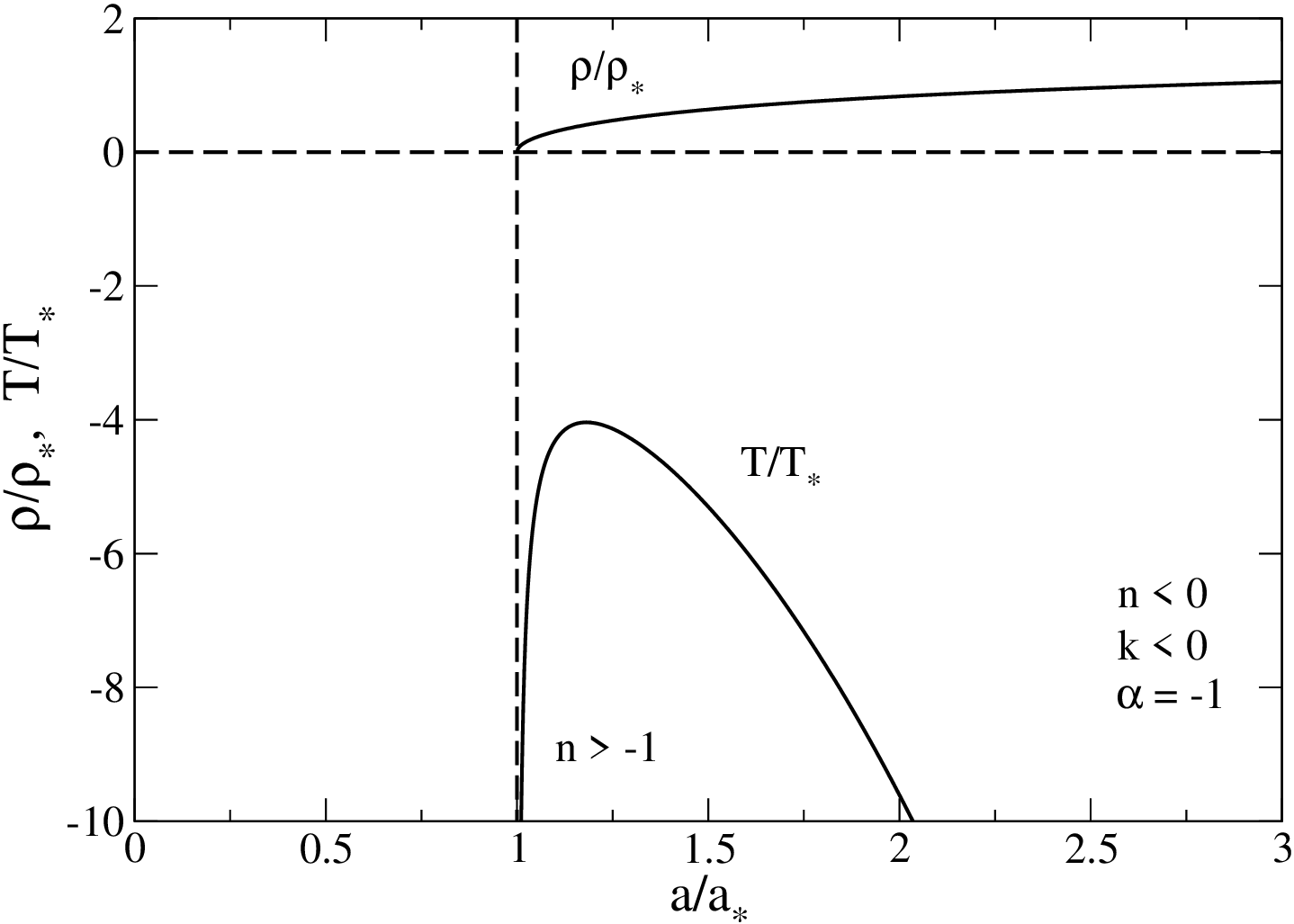}
\caption{Evolution of the density and temperature as a function of the scale factor. We have taken $n=-1/2$.}
\label{newannexedensityNEG}
\end{center}
\end{figure}

The thermodynamical equation (\ref{a3}) can be integrated into
\begin{eqnarray}
\label{eosgm2b}
T=-T_* \left (\frac{\rho}{\rho_*}\right )^{(n+1)/n} e^{3(\rho_*/\rho)^{1/n}},
\end{eqnarray}
where $T_*>0$ is a constant of integration. Combined with Eq. (\ref{eosgm14}), we obtain
\begin{eqnarray}
\label{eosgm3b}
T= -\frac{T_*}{\ln(a/a_*)^{n+1}}\left (\frac{a}{a_*}\right )^3.
\end{eqnarray}
When $a\rightarrow a_*$, $T\rightarrow 0$ for $n<-1$ and $T\rightarrow -\infty$ for $n>-1$. When $a\rightarrow +\infty$, $T\rightarrow -\infty$. For $n<-1$, the temperature reaches its maximum at
\begin{equation}
\label{eosgm19b}
\frac{\rho_{e}}{\rho_*}=\left (\frac{3}{n+1}\right )^n,\qquad \frac{a_{e}}{a_*}=e^{(n+1)/3},
\end{equation}
\begin{equation}
\label{eosgm4bb} \frac{T_e}{T_*}=-\left (\frac{3}{n+1}\right )^{n+1}e^{n+1}.
\end{equation}
The evolution of the density and temperature as a function of the scale factor is represented in Fig. \ref{newannexedensityNEG}.

The equation of state can be written as $p=w\rho c^2$ with
\begin{equation}
\label{eosgm17}
w=-1+\frac{n}{3}\left (\frac{\rho}{\rho_*}\right )^{1/n}.
\end{equation}
When $a\rightarrow a_*$, $w\rightarrow -\infty$; when  $a\rightarrow +\infty$, $w\rightarrow -1$.

The deceleration parameter is given by Eqs. (I-77) and (I-78). Together with Eq. (\ref{eosgm17}), we obtain
\begin{equation}
\label{eosgm18}
q=-1+\frac{n}{2}\left (\frac{\rho}{\rho_*}\right )^{1/n}.
\end{equation}
When $a\rightarrow a_*$, $q\rightarrow -\infty$; when $a\rightarrow +\infty$, $q\rightarrow -1$.

\begin{figure}[!h]
\begin{center}
\includegraphics[clip,scale=0.3]{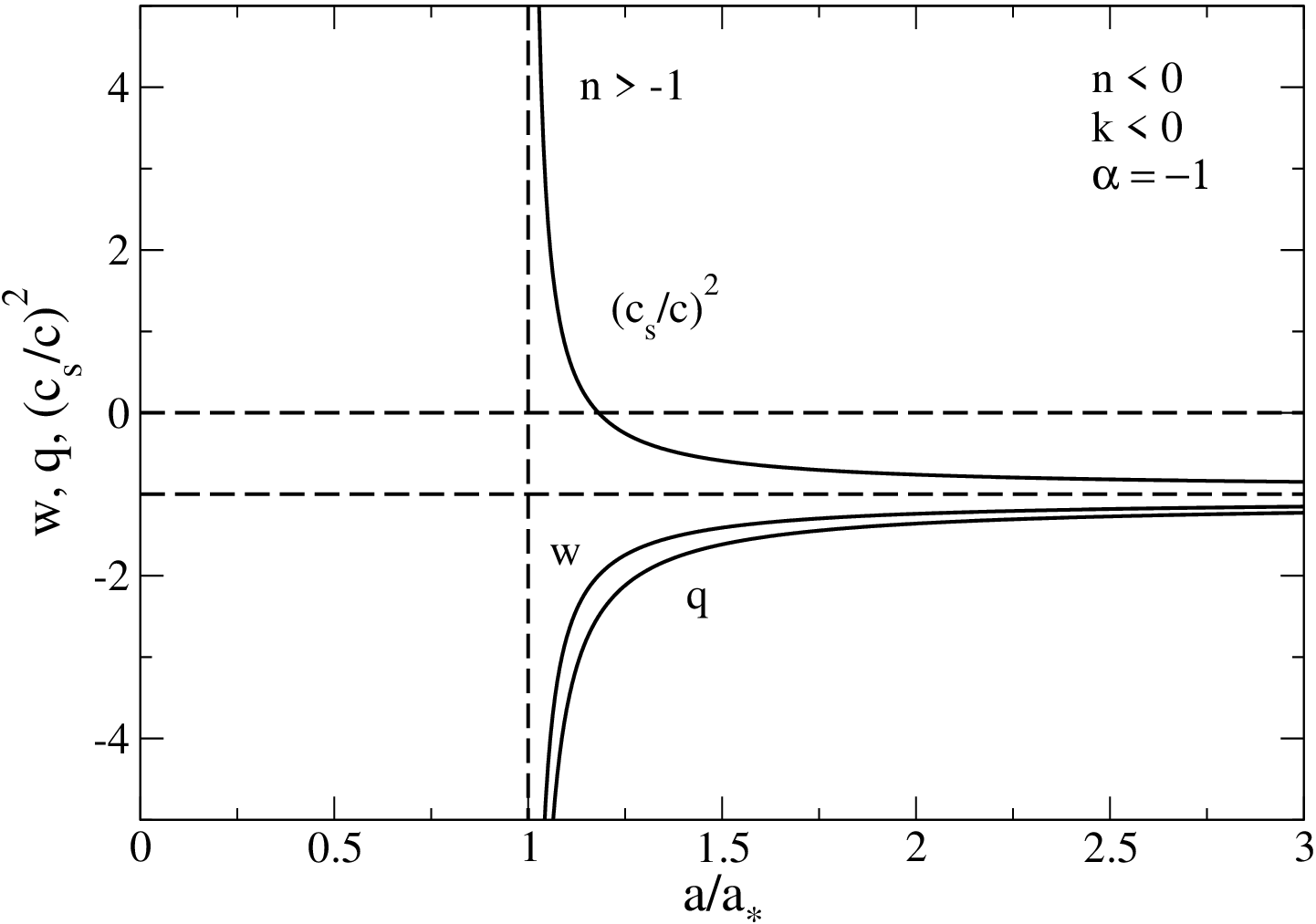}
\caption{Evolution of $w$, $q$, and $(c_s/c)^2$ as a function of the scale factor $a$. We have taken $n=-1$. }
\label{newannexewNEG}
\end{center}
\end{figure}

The velocity of sound is given by
\begin{equation}
\label{eosgm19}
\frac{c_s^2}{c^2}=-1+\frac{n+1}{3}\left (\frac{\rho}{\rho_*}\right )^{1/n}.
\end{equation}
We have to distinguish several cases. We first assume $n<-1$. When $a\rightarrow a_*$, $(c_s/c)^2\rightarrow -\infty$; when $a\rightarrow +\infty$, $(c_s/c)^2\rightarrow -1$.  The velocity of sound is always imaginary. We now assume $n>-1$. When $a\rightarrow a_*$, $(c_s/c)^2\rightarrow +\infty$; when $a\rightarrow +\infty$, $(c_s/c)^2\rightarrow -1$. The velocity of sound vanishes at the point (\ref{eosgm19b}) at which the temperature is maximum. At that point, the pressure is maximum with value
\begin{equation}
\label{gdoe}
\frac{p_{e}}{\rho_*c^2}=-\frac{3^n}{(n+1)^{n+1}}.
\end{equation}
The velocity of sound is real for $a<a_e$ and imaginary for $a>a_e$. On the other hand, the velocity of sound is equal to the speed of light at
\begin{equation}
\label{eosgm19c}
\frac{\rho_{s}}{\rho_*}=\left (\frac{6}{n+1}\right )^n,\qquad \frac{a_{s}}{a_*}=e^{(n+1)/6}.
\end{equation}
The velocity of sound is larger than the speed of light when $a<a_s$ and smaller when $a>a_s$. The evolution of $w$, $q$, and $(c_s/c)^2$ as a function of the scale factor $a$ is represented in Fig. \ref{newannexewNEG}.

 \begin{figure}[!h]
\begin{center}
\includegraphics[clip,scale=0.3]{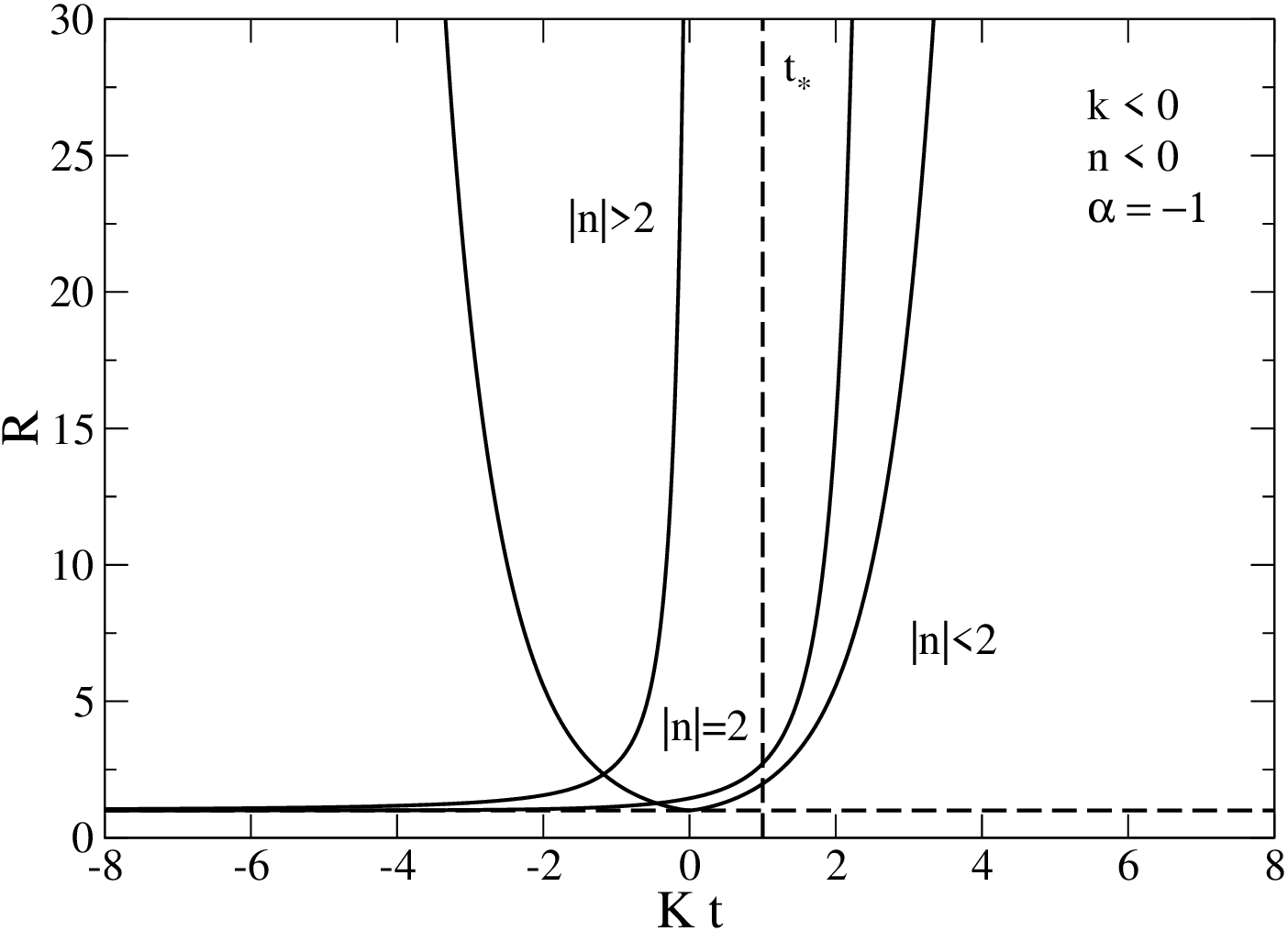}
\caption{Evolution of the radius with time for $n<0$. We have taken $n=-3$, $n=-2$ and $n=-1/2$. For $n\le -2$, we have taken $Kt_{*}=1$. For $n>-2$, the universe is bouncing at $t=0$ (see text for details). }
\label{annexeKnegNneg}
\end{center}
\end{figure}

Setting $R=a/a_*$, the Friedmann equation (\ref{basic2b}) can be written
\begin{eqnarray}
\label{eosgm20}
\dot R=\frac{\epsilon K R}{(\ln R)^{n/2}},
\end{eqnarray}
where $K=(8\pi G\rho_*/3)^{1/2}$ and $\epsilon=\pm 1$. We must distinguish three cases.

(i) For $n<-2$,
\begin{eqnarray}
\label{eosgm22}
R(t)=e^{\left \lbrack \frac{|n|-2}{2} K(t_{*}-t)\right \rbrack^{-2/(|n|-2)}},
\end{eqnarray}
\begin{eqnarray}
\label{marre7}
\frac{\rho(t)}{\rho_*}=\left \lbrack \frac{|n|-2}{2} K(t_{*}-t)\right \rbrack^{-2|n|/(|n|-2)}.
\end{eqnarray}
The universe starts from $t=-\infty$ with a finite radius $R=1$, a vanishing density, and a vanishing pressure (past peculiarity). It undergoes a future singularity of type I (Big Rip): At $t=t_{*}$, the scale factor, the density and the pressure are infinite. The divergence of the density is algebraic.

(ii) For $n=-2$,
\begin{eqnarray}
\label{eosgm23}
R(t)=e^{e^{K(t-t_*)}},
\end{eqnarray}
\begin{eqnarray}
\label{marre8}
\frac{\rho(t)}{\rho_*}=e^{2K(t-t_*)}.
\end{eqnarray}
The universe starts from $t=-\infty$ with a finite radius $R=1$, a vanishing density, and a vanishing pressure (past peculiarity). There is no future singularity. For $t\rightarrow +\infty$, the radius and the density tend to $+\infty$ and the pressure to $-\infty$ (Little Rip). The density increases exponentially rapidly.

(iii) For $n>-2$,
\begin{eqnarray}
\label{eosgm21}
R(t)=e^{\left (\frac{2-|n|}{2} Kt\right )^{2/(2-|n|)}}.
\end{eqnarray}
\begin{eqnarray}
\label{marre6}
\frac{\rho(t)}{\rho_*}=\left (\frac{2-|n|}{2} Kt\right )^{2|n|/(2-|n|)}.
\end{eqnarray}
The universe starts at $t=0$ with a finite radius $R=1$ and a vanishing density (past peculiarity).
At $t=0$, the pressure vanishes for $-2<n<-1$, is finite for $n=-1$ and tends to $-\infty$ for $n>-1$. In this last case, there is a past singularity of type II. On the other hand, there is no future singularity: For $t\rightarrow +\infty$, the radius and the density tend to $+\infty$ and the pressure to $-\infty$ (Little Rip). Actually, we can extend the solution to $t<0$ (except, maybe, in the case $n>-1$ where the pressure diverges at $t=0$). This describes a phase of contraction of the universe, corresponding to the solution of Eq. (\ref{eosgm20}) with $\epsilon=-1$. Therefore, we obtain a model of bouncing phantom universe that collapses for $t<0$ (with decreasing density), disappears at $t=0$ (the density vanishes), and expands for $t>0$ (with increasing density).

The evolution of the scale factor in these different cases is represented in Fig. \ref{annexeKnegNneg}.

\section{Summary of all the possible cases}
\label{sec_summary}

The study of the polytropic equation of state (\ref{intro1}) in
cosmology is very rich. This is also the case for the study of polytropic distributions in stellar structure
\cite{chandra} and in other areas of
physics and biology \cite{cs,cc,nfp}. In this Appendix, we summarize
all the results obtained in our series of papers and analyze the
different singularities in terms of the classification of
\cite{classification}.

\subsection{The case $-1<\alpha\le 1$}

$\bullet$ In papers I and II, we have studied the case $w\ge -1$. We have obtained the following results:

(i) For $n>0$ and $k<0$, the universe undergoes an early inflation. It starts from $t=-\infty$ with a vanishing radius $a=0$ and a finite density $\rho_{*}$. Its radius increases indefinitely in time while its density decreases. There is no singularity.

(ii) For $n>0$ and $k>0$, the universe exhibits a past singularity of type III. It starts at $t=0$ with a finite radius $a_*$ and an infinite density $\rho=+\infty$. For $t>0$, its radius increases indefinitely in  time while its density decreases. There is no future singularity.

(iii) For $n<0$ and $k<0$, the universe exhibits a  Big Bang singularity. It starts at $t=0$ with a vanishing radius $a=0$ and an infinite density $\rho=+\infty$. The universe also undergoes a late inflation. Its radius increases to $+\infty$ as  $t\rightarrow +\infty$ while its density decreases to a finite value $\rho_{*}$. There is no future singularity.

(iv) For $n<0$ and $k>0$, the universe exhibits a Big Bang singularity. It starts at $t=0$ with a vanishing radius $a=0$ and an infinite density $\rho=+\infty$. The universe also exhibits a future peculiarity. Its radius increases to a finite value $a_*$ while its density decreases to zero $\rho=0$ (the universe ``disappears''). For $n\le -2$, this peculiarity is reached in infinite time. For $n>-2$, this peculiarity is reaches in a finite time $t_*$  (for $n<-1$, there is a future singularity of type II because the pressure diverges at $t=t_*$). For $t_*<t<2t_*$, the radius decreases to zero while the density increases to $+\infty$. This leads to a Big Crunch singularity. These phases of expansion and contraction continue periodically (cyclic universe).

$\bullet$ In this paper, we have studied the case $w<-1$ (requiring $k<0$) corresponding to a phantom universe. We have obtained the following results:

(i) For $n>0$, the universe undergoes an early inflation. It starts from $t=-\infty$ with a vanishing radius $a=0$ and a finite density $\rho_{*}$. The universe also undergoes a future singularity of type III: At a finite time $t_*$, its radius tends to a finite value $a_*$ while its density diverges $\rho\rightarrow +\infty$.

(ii) For $n<0$, the universe exhibits a past peculiarity. Its radius
starts from a finite value $a_*$ while its density vanishes
$\rho=0$. For $n\le -2$, this peculiarity occurs in the infinite
past. For $n>-2$, this peculiarity occurs at $t=0$ (for $n>-1$, there
is a past singularity of type II because the pressure diverges at
$t=0$).  The universe also undergoes a late inflation. Its radius
increases to $+\infty$ as $t\rightarrow +\infty$ while its density
increases to a finite value $\rho_{*}$. There is no future
singularity. Actually, the solutions with $n>-2$ can be continued
symmetrically for $t<0$ leading to models of bouncing phantom universe.

\subsection{The case $\alpha=-1$}

$\bullet$ In papers I and II we have studied the case $w>-1$ (requiring $k>0$). We have obtained the following results:

(i) For $n>0$, the universe exhibits a past singularity of type III. It starts at $t=0$ with a finite radius $a_*$ and an infinite density $\rho=+\infty$. For $t>0$, its radius increases indefinitely in  time while its density decreases. There is no future singularity.

(ii) For $n<-2$, the universe exhibits a Big Bang singularity. It starts at $t=0$ with a vanishing radius $a=0$ and an infinite density $\rho=+\infty$. The universe also exhibits a future peculiarity. Its radius increases to a finite value $a_*$ while its density decreases to zero $\rho=0$. This peculiarity is reached algebraically rapidly in infinite time.

(iii) For $n=-2$, the universe starts from $t=-\infty$ with a vanishing radius $a=0$ and an infinite density $\rho=+\infty$. The universe exhibits a future peculiarity. Its radius increases to a finite value $a_*$ while its density decreases to zero $\rho=0$. This peculiarity is reached exponentially rapidly in infinite time.

(iv) For $-2<n<0$, the universe starts from $t=-\infty$ with a
vanishing radius $a=0$ and an infinite density $\rho=+\infty$. The
universe exhibits a future peculiarity. Its radius increases to a
finite value $a_*$ while its density decreases to zero $\rho=0$ (the
universe ``disappears''). This peculiarity is reached in a finite time
$t_*$ (for $-1<n<0$, there is a future singularity of type II because
the pressure diverges at $t=t_*$). For $t>t_*$, the radius decreases
to zero while the density increases to $+\infty$. This takes place in
infinite time.

$\bullet$ In this paper, we have studied the case $w<-1$ (requiring $k<0$) corresponding to a phantom universe. We have obtained the following results (see also \cite{stefancic,classification}):

(i) For $n>0$ the universe starts from $t=-\infty$ with a vanishing radius $a=0$ and a vanishing density $\rho=0$. This corresponds to a  generalized past peculiarity. The universe also exhibits a future singularity of type III: At a finite time $t_*$, its radius tends to a finite value $a_*$ while its density diverges $\rho\rightarrow +\infty$.

(ii) For $n<-2$, the universe exhibits a past peculiarity. It starts from $t=-\infty$ with a finite radius $a_*$ and a vanishing density $\rho=0$. It also exhibits a future singularity of type I (Big Rip): At a finite time $t_*$, its radius and density are infinite. This singularity is reached algebraically rapidly.

(iii) For $n=-2$, the universe exhibits a past peculiarity. It starts from $t=-\infty$ with a finite radius $a_*$ and a vanishing density $\rho=0$. Then, the radius and the density increase indefinitely (Little Rip).

(iv) For $-2<n<0$, the universe exhibits a past peculiarity: It starts
at $t=0$ with a finite radius $a_*$ and a vanishing density $\rho=0$
(for $-1<n<0$, there is a past singularity of type II because the
pressure diverges at $t=0$). Then, the radius and the density increase
indefinitely (Little Rip). Actually, the solution can be continued
symmetrically for $t<0$. This leads to a model of bouncing phantom universe.

\end{document}